\documentclass[journal,10pt,]{IEEEtran}
\usepackage[hidelinks]{hyperref}
\usepackage[utf8]{inputenc}
\usepackage[T1]{fontenc}
\usepackage{placeins}
\usepackage{microtype}
\usepackage[dvips]{graphicx}
\usepackage{caption}
\graphicspath{ {images/} }
\usepackage{xcolor}
\usepackage{times}
\usepackage{amsmath}
\usepackage{float}
\usepackage{epstopdf}
\usepackage{amssymb}
\usepackage{blindtext}
\usepackage{enumitem}
\usepackage{xcolor}
\usepackage{amsmath}
\usepackage{epstopdf}
\usepackage{array}
\usepackage{multirow}
\usepackage{algorithm}
\usepackage{cite}
\usepackage{tikz}
\usetikzlibrary{shapes}
\usetikzlibrary{plotmarks}
\usetikzlibrary{spy}
\usepackage{import}
\usepackage{pgfplots}
\usepackage{mathtools}
\usepackage{multirow} 
\usepackage{adjustbox}
\usetikzlibrary{spy}
\usepackage[utf8]{inputenc}
\usepackage{tikz}
\usetikzlibrary{matrix,decorations.pathreplacing}
\usetikzlibrary{shapes}
\usepackage[utf8]{inputenc}
\usepackage{amsmath}
\usepackage{amssymb}
\usepackage{amsthm}
\usetikzlibrary{plotmarks}
\usepackage{pgfplots}
\usetikzlibrary{spy}
 \usepackage{pstricks-add}
 \usepackage{epsfig}
 \usepackage{pst-grad} 
 \usepackage{pst-plot} 
 \usepackage[space]{grffile} 
 \usepackage{etoolbox}

 \definecolor{darkblue}{rgb}{0.0, 0.0, 0.55}
 \usepackage{tikz}
 \usepackage{pgfplots}
\usepackage[a4paper,margin=2cm]{geometry}
\usepackage{forest}
\usetikzlibrary{trees,positioning,shapes,shadows,arrows.meta}

\usepackage[labelformat=simple]{subcaption}

\usepgfplotslibrary{external}
\usepackage{algpseudocode}
\usepackage{tabularx}
\usepackage{booktabs}
\usepackage{array}
\usepackage[font=small]{caption}

\usepackage{etoolbox}
\AtBeginEnvironment{algorithm}{\algoequations}
\AtEndEnvironment{algorithm}{\restoreequations}
\newcounter{algosavedequation}
\newcommand{\algoequations}{%
  \setcounter{algosavedequation}{\value{equation}}%
  \setcounter{equation}{0}%
  \renewcommand{\theequation}{A.\arabic{equation}}%
}
\newcommand{\innerproduct}[2]{\langle #1, #2 \rangle}
\newcommand{\restoreequations}{%
  \setcounter{equation}{\value{algosavedequation}}%
}

\DeclareMathOperator*{\argmin}{argmin}
\makeindex 
\ifCLASSINFOpdf
  
\else

\fi

\definecolor{mygr}{HTML}{e6e6e6}
\definecolor{bblue}{HTML}{4F81BD}
\definecolor{rred}{HTML}{C0504D}
\definecolor{ggreen}{HTML}{9BBB59}
\definecolor{ppurple}{HTML}{9F4C7C}
\definecolor{antiquebrass}{rgb}{0.8, 0.58, 0.46}
\definecolor{mycolor}{rgb}{1,0.4,0.9}

\newtheorem{theorem}{Theorem}

\newtheorem{lemma}{Lemma}
\makeatletter 
\pretocmd\@bibitem{\color{black}\csname keycolor#1\endcsname}{}{\fail}
\newcommand\citecolor[1]{\@namedef{keycolor#1}{\color{blue}}}
\makeatother
\citecolor{Liu_Lei}
\citecolor{Ma_Ping}
\citecolor{Chen_Liu}
\citecolor{Singer_Andrew}
\citecolor{Hayakawa}
\citecolor{3gpp.23.003}
\citecolor{minsum}
\citecolor{donoho2009}
\begin{document}

\tikzset{new spy style/.style={spy scope={%
			magnification=2.8,
			size=1.25cm,
			connect spies,
			every spy on node/.style={
				rectangle,
				draw,
			},
			every spy in node/.style={
				draw,
				rectangle,
			}
		}
	}
}

\newcolumntype{L}[1]{>{\raggedright\let\newline\\\arraybackslash\hspace{0pt}}m{#1}}
\newcolumntype{C}[1]{>{\centering\let\newline\\\arraybackslash\hspace{0pt}}m{#1}}
\newcolumntype{R}[1]{>{\raggedleft\let\newline\\\arraybackslash\hspace{0pt}}m{#1}}
\pgfdeclarelayer{background}
\pgfdeclarelayer{foreground}
\pgfsetlayers{background,main,foreground}
\newcommand{\oeq}{\mathrel{\text{\sqbox{$=$}}}}
\setlength{\textfloatsep}{0.1cm}
\setlength{\floatsep}{0.1cm}
\tikzstyle{int}=[draw, fill=white!20, minimum size=2em]
\tikzstyle{init} = [pin edge={to-,thin,black}]

\title{Robust Prior Information-Aided ADMM for Multi-User Detection in Codebook-Based Grant-Free NOMA Under Dynamic Scenarios }
\author{Vinjamoori Vikas, Kuntal Deka, Sanjeev Sharma, A. Rajesh}

\maketitle  

\begin{abstract}
Code-domain non-orthogonal multiple access (CD-NOMA) systems offer key benefits such as high spectral efficiency, low latency, high reliability, and massive connectivity. NOMA's ability to handle overloading allows multiple devices to share a single resource element (RE) for data transmission. In CD-NOMA, different users are assigned distinct codewords, which are leveraged during multi-user detection (MUD). Codebook-based NOMA systems outperform spread-sequence (SS)-based NOMA due to the coding gains provided by the codebooks. Sparse code multiple access (SCMA) and dense code multiple access (DCMA) are two prominent examples of such systems. Additionally, NOMA is seen as a crucial technology for enabling grant-free access, especially in massive machine-type communications (mMTC).
One of the main challenges in deploying grant-free NOMA systems is accurately detecting both user activity and transmitted data, particularly when user activity fluctuates dynamically across the transmission frame. This paper introduces codebook-based grant-free NOMA systems modeled using a block sparsity signal structure. The joint activity and data detection (JADD) problem in these systems is formulated as group LASSO and sparse group LASSO block compressive sensing (BCS) problems. To address these, a robust prior information-aided alternating direction method of multipliers (ADMM) algorithm is proposed. Extensive numerical experiments and theoretical analysis show the efficiency of the proposed algorithm, making it a suitable solution for mMTC networks.

\end{abstract}

\IEEEpeerreviewmaketitle

\section{Introduction}
The non-orthogonal multiple access (NOMA) systems are well investigated to address several challenges of 6G wireless networks~~\cite{NOMA_survey}. The challenges include high spectral efficiency, low latency, massive connectivity, and high reliability, to name a few. The two operating domains of NOMA, power domain NOMA (PD-NOMA) and code-domain NOMA (CD-NOMA), are extensively documented in the literature. CD-NOMA offers various benefits over PD-NOMA in terms of error rate performance, flexible system design, etc. ~\cite{NOMA_survey}. CD-NOMA enjoys additional coding gain provided by the multi-dimensional constellations structure of codebook ~\cite{Nikopour, Liu}.

 Grant-free NOMA is a critical enabling technology for grant-free access in the NOMA-aided networks~\cite{GF_NOMA_survey}. 
 In this work, NOMA is often used to refer to CD-NOMA.   NOMA enables the simultaneous transmission of multi-user data on a single resource element (RE). Thus, data overloading is allowed in NOMA. Hence, NOMA supports massive connectivity with a limited number of resource elements (REs). On the other hand, orthogonal multiple access (OMA) allocates each RE exclusively to a single user for transmission. OMA can schedule grant-based access, where each user pre-allocates an orthogonal channel. This approach is prevalent in human-type communication (HTC) or human-to-human communications, which typically involve fewer devices with large amounts of data.
 However, this approach introduces additional overhead and latency, which is often negligible compared to the large amount of data transmission.
 This assumption can't hold for massive machine-type communications (mMTC) networks.
 
 6G wireless networks hold promise for mMTC, where a massive number of devices are connected to the network. Moreover, these devices sporadically transmit short-packet data. The conventional OMA scheme becomes a potential bottleneck for the mMTC. Indeed, OMA is not a feasible idea for the massive connectivity. The primary challenge of the 6G network is to provide massive connectivity with limited resources. 
 Furthermore, ultra-reliable low-latency communication (URLLC) is the essential requirement of a 6G wireless network. 
 Conventional grant-based access in HTC is no longer possible in mMTC. The overloading nature of NOMA supports massive connectivity, and grant-free access enables URLLC. Therefore, grant-free NOMA  enjoys massive connectivity and URLLC simultaneously~\cite{GF_NOMA_survey}.

 
 In mMTC, short packet transmissions, massive devices, and sporadic traffic, the orthogonal resource allocation and pre-scheduling are not feasible. Moreover, the grant-based scheme doesn't enjoy the idea of massive devices with limited resources. A NOMA scheme is a potential solution for implementing mMTC networks. It supports massive connectivity with limited resources by overloading the devices on each resource. Further, NOMA exploits overloading to provide grant-free access to mMTC. Devices can transmit the data anytime without pre-scheduling and grant from the base station (BS).
 
 In mMTC, only 10\%  of the total devices in the network are active even in the busy hour due to its sporadic traffic~\cite{wang_dai}. The NOMA has been considered a potential enabler for providing grant-free access to the massive devices in the network. Hence, each device can access the network without any scheduling. 
 The devices randomly arrive and depart from the network and transmit the data whenever required.
 The receiver, or a BS in the UL scenario, does not have any information about the users' activity at any point in time. Receivers need to have an additional mechanism to identify the active users in the network.
 The main challenge of Grant-free NOMA systems lies in simultaneously detecting the active users and their data. The Joint activity and data detection (JADD) with acceptable performance and complexity trade-off is an active area of research. The transmitted signal structure influences the JADD of grant-free NOMA. The transmitted signal structure leads to two variants in CD-NOMA, as discussed below.

Based on the code structure, two variants of CD-NOMA have been extensively studied, as shown in Fig.~\ref{fig1_CD}. The sequence-based NOMA performs mapping and spreading operations separately in the encoding. Sequence-based NOMA, namely, Low-density signature (LDS)  and multi-user shared access (MUSA)~\cite{MUSA}, exist. LDS uses sparse spreading sequences, and MUSA uses dense spread sequences—the spreading sequences used in LDS and MUSA are relatively simple structures and easy to design.    
Codebook-based NOMA (CB-based NOMA) merges mapping and spreading operations of sequence-based NOMA~\cite{Nikopour}. 
In this approach,  data is mapped directly to codewords during the encoding process. The codewords are drawn from the pre-designed codebooks. Unlike simple, complex sequences, the codebooks are designed from multi-dimensional constellations rather than simple, complex sequences, offering additional shaping gain. 
Two different types of codebook structures are available in the literature. The first is sparse code multiple access (SCMA), and the second is dense code multiple access (DCMA). SCMA utilizes sparse codebooks to reduce the complexity of the detection. The exponentially complex message-passing algorithm (MPA) is extensively used for SCMA detection. SCMA suffers from limited diversity order. DCMA is another prominent CB-based NOMA technique. It uses dense codebooks for data encoding. DCMA fully utilizes the dimensions of the codebook. Thus, it enjoys full diversity gains. Generalized sphere decoder (GSD) is used for DCMA detection. It shows high computational complexity because the GSD works based on the tree search method. 
Later, a competent algorithm based on the ADMM optimization method is proposed for CB-based NOMA systems~\cite{ADMM_vik}. The authors proved it is an efficient detector for SCMA and DCMA with acceptable complexity.  However, all current multi-user detection algorithms for CB-based NOMA systems are solely designed for data detection. They are not tailored for grant-free NOMA systems and do not support joint activity and data detection (JADD).

   \tikzset{
   	basic/.style  = {draw, text width=2cm, drop shadow, font=\sffamily, rectangle},
   	root/.style   = {basic, rounded corners=2pt, thin, align=center, fill=white},
   	level-2/.style = {basic, rounded corners=6pt, thin,align=center, fill=white, text width=3cm},
   	level-3/.style = {basic, thin, align=center, fill=white, text width=2.8cm}
   }
   \tikzset{
   	basic/.style  = {draw, text width=2cm, drop shadow, font=\sffamily, rectangle},
   	root/.style   = {basic, rounded corners=2pt, thin, align=center, fill=white},
   	level-2/.style = {basic, rounded corners=6pt, thin,align=center, fill=white, text width=3cm},
   	level-3/.style = {basic, thin, align=center, fill=white, text width=1.8cm}
   }
   \begin{figure}
   	\centering
   	\begin{tikzpicture}[
   		level 1/.style={sibling distance=12em, level distance=5em},
   		edge from parent/.style={->,solid,black,thick,sloped,draw}, 
   		edge from parent path={(\tikzparentnode.south) -- (\tikzchildnode.north)},
   		>=latex, node distance=1.2cm, edge from parent fork down]
   		
   		\node[root] {\textbf{CD-NOMA}}
   		child {node[level-2] (c1) {\textbf{Sequence based}}}
   		child {node[level-2] (c2) {\textbf{Codebook based}}};
   		
   		\begin{scope}[every node/.style={level-3}]
   			\node [below of = c1, xshift=10pt] (c11) {LDS};
   			\node [below of = c11] (c12) {MUSA};
   			
   			\node [below of = c2, xshift=10pt] (c21) {SCMA};
   			\node [below of = c21] (c22) {DCMA};
   			
   			%
   		\end{scope}
   		
   		\foreach \value in {1,2}
   		\draw[->] (c1.195) |- (c1\value.west);
   		
   		\foreach \value in {1,2}
   		\draw[->] (c2.195) |- (c2\value.west);
   		
   		%
   	\end{tikzpicture}
   	\caption{Classification of CD-NOMA}
   	\label{fig1_CD}
   \end{figure}
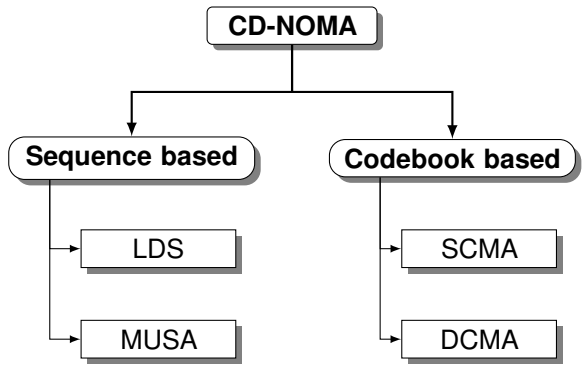

   Extensive research has been conducted on the multi-user detection of grant-free NOMA systems for JADD.
  The existing work on JADD focuses mainly on spread-sequence-based grant-free CD-NOMA systems in mMTC networks. Due to the sparse nature of the input signaling due to sporadic traffic, the JADD of grant-free NOMA is formulated as a sparse recovery problem. Standard compressive sensing (CS) based multi-user detection is a practical approach to solve the JADD problem. Numerous CS-based JADD algorithms have been proposed in the literature, and they can be broadly classified into three categories: greedy-based algorithms, Bayesian inference or message-passing algorithms, and convex optimization-based algorithms. Each of these approaches comes with its own set of strengths and limitations.
   
  The low complex greedy algorithms popularly known as orthogonal matching pursuit (OMP)~\cite{Wang_OMP} and subspace pursuit (SP)~\cite{Dai_SP} algorithms are extensively used for JADD. These algorithms consider sparsity level ($S_{\rm l}$) as prior information.  
 The CS-based OMP-aided MUD is proposed for downlink (DL) CDMA systems \cite{Shim_CS_CDMA}. The structured sparsity-based MUD algorithm known as structured iterative support detection (SISD) is proposed for the JADD of grant-free NOMA systems~\cite{wang_dai}. Structured sparsity is a restricted assumption of the activity of the devices. Later, research has been conducted on dynamic activity scenarios. A dynamic system model is considered, and a dynamic CS (DCS)--based MUD  is proposed based on SP algorithm~\cite{wang_dynamic}.
 Further, the temporal correlation of the active user sets between two adjacent time slots is exploited to improve the performance of the detector. DCS-based MUD considers the known user sparsity level unrealistic. It blindly utilizes the support set from the previous time slot, disregarding that only a subset of the current time slot's support set overlaps with the previous one. 
 The prior-information aided DCS algorithm is proposed for JADD of dynamic grant-free NOMA system~\cite{prior}. It exploits the temporal correlation by considering the quality of information that exists between the consecutive time slots. Thus, it achieves significant performance gains compared with the earlier works. A sparse recovery approach, where the non-zeros occur in clusters in the sparse signal, is known as block sparse recovery or block compressive sensing (BCS) problems~\cite{Elder_block}. A block sparsity system model is proposed for continuous data transmission in UL grant-free NOMA systems~\cite{du_block}. BCS-based MUD is proposed for JADD and achieves further performance gains compared to the structured sparsity model. However, this model can not be adaptable to dynamic scenarios. All the above detectors are built on greedy-based algorithms, including OMP and SP, etc. These algorithms computes the heuristic local support set in each iteration and requires to performs computationally expensive inverse operation in each iteration. It does not guarantee the global optimum, especially when the signal is not very sparse or the noise effect is more significant. Further, it performs well for incoherent measurement matrices, and the performance degrades for the coherent matrices.

   The approximate message passing (AMP) is another approach based on a message passing algorithm (MPA) to solve the JADD problem in grant-free NOMA~\cite{Wei_AMP_NOMA}.
 AMP relies on the assumption that the entries of the channel matrix are independent and identically distributed (IID) to carry out state evolution (SE) analysis. It performs poorly when the channel matrix is not IID~\cite{OAMP}. An orthogonal AMP-based MUD is proposed for JADD of grant-free NOMA systems~\cite{AMP_2}. Both of these works exploit the structured sparsity model for performance enhancement. However, the structured sparsity assumption is often unrealistic and not well suited for mMTC networks. Further, they ignores the challenges of dynamic user activity scenarios.

   The convex optimization-based approaches have also been well studied in CS, popularly known as the Least Absolute Shrinkage and Selection Operator (LASSO)~\cite{LASSO}. Despite this, the alternative direction method of multipliers (ADMM) is an efficient algorithm to solve the convex or non-convex problems\cite{boyd}. The decomposition and parallel computing properties help to solve complex problems. Indeed, ADMM is an efficient approach to solving problems with large-scale data. Strong theoretical convergence properties made ADMM more robust in a wide range of applications. Later, ADMM is adapted to solve the LASSO problem~\cite{boyd}. 
  The LASSO-based MUD is formulated for JADD of grant-free NOMA system ~\cite{Cirik}. The ADMM algorithm is used to solve the formulated LASSO- multi-user detection problem. The joint processing of the structured sparsity model in this work improves the performance of the ADMM detector. Further, the proposed ADMM algorithm is adapted to dynamic scenarios. However, blindly exploiting the temporal correlation between consecutive time slots in a dynamic model leads to performance degradation. There is still room for further improvement in the performance of ADMM-based MUD in dynamic scenarios.

The comprehensive survey outlined above introduces various models for grant-free NOMA systems and proposes new algorithms to tackle the multi-user detection challenge for JADD. However, all of these studies focus on JADD for grant-free spread-sequence-based NOMA systems. To the best of the author's knowledge, no prior work has explored the design of a CB-based NOMA system for grant-free scenarios or developed a detection algorithm to address the JADD problem in such systems. As discussed earlier, CB-based grant-free NOMA offers several advantages over sequence-based grant-free NOMA. Additionally, the inherent block-sparsity structure in CB-based grant-free NOMA enhances the robustness of sparse signal reconstruction.
  
  The research on the codebook design for NOMA systems is well-developed. This idea inspired the authors to develop new models to solve the problems in the practical implementation of mMTC networks by leveraging the CB-based grant-free NOMA systems. The two block-CS (BCS) problems are formulated to address the JADD problem of introduced block-sparsity models. The two BCS problems include sparse group LASSO and group LASSO for SCMA and DCMA, respectively. A distributed optimization-aided ADMM algorithm is designed to solve the formulated BCS problems.    
  This work introduces a novel frame-wise dynamic block sparsity model for a grant-free NOMA system. Further, we have introduced a comprehensive treatment of a dynamic scenario to address major critical issues in mMTC networks in practice. A robust prior information-aided ADMM algorithm is developed for JADD in dynamic scenarios. It is different from the existing prior information-aided SP algorithm~\cite{prior}.   This work assumes the quality parameter of the prior information known to the receiver, which needs to be more accurate. Obtaining accurate quality of prior information without computing the active user support is challenging; setting an inaccurate quality parameter in~\cite{prior} results in detrimental effects in JADD. To resolve this, a systematic two-step computation process is proposed to leverage the temporal correlation that exists in the dynamic scenarios. Accurate quality of prior information is obtained in Step 1 and is incorporated to improve the quality of JADD in Step 2. 
   The detailed contributions of this work listed below:\\
 $\bullet$ \textbf{\textit{Block CS (BCS) models for codebook based CD-NOMA systems:}}\\
 The grant-free codebook based CD-NOMA (SCMA and DCMA) systems possesses inherent block sparsity structure in its input signaling. The proposed system models enjoys the block sparsity structure of CD-NOMA codewords. Indeed the JADD problem exploits the BCS approaches to enhance the performance of multi-user detection. Block sparsity based frame-wise dynamic system model is proposed and extensive numerical experiments have been discussed .\\
 $\bullet$ \textbf{\textit{JADD problem formulation:}}\\
The high computationally complex detection problem of SCMA and DCMA is solved by using 
 a convex optimization framework.
 Thus, the multi-user detection problem for JADD of SCMA and DCMA simplified to block sparse recovery problem.
   The JADD problem of sparse and dense codebook based CD-NOMA systems  is formulated as two closely related BCS approaches. The efficient sparse group LASSO and group LASSO BCS approaches are proposed to solve the detection problem of SCMA and DCMA, respectively. 
 
 $\bullet$ \textbf{\textit{JADD problem solved using the  ADMM algorithm:}}\\
A parallel computing aided ADMM based detection algorithm is proposed to solve the formulated JADD problem. The distributive nature of ADMM makes it suitable for large-scale CD-NOMA systems which expected to exists in mMTC networks. Hence, the proposed detector is scalable unlike the message passing detector (MPA) and generalized sphere decoder (GSD) available for both SCMA and DCMA  systems, respectively.  \\
$\bullet$ \textbf{\textit{Block-sparsity based dynamic compressive sensing (DCS) approach:}}\\
The random arrivals and departures of nodes in mMTC network makes the JADD further challenging. The  dynamic system model is further enhanced by introducing practical use cases, which are derived from more realistic assumptions. Block-sparsity aided dynamic compressive sensing (DCS) based ADMM algorithm is proposed.  Thanks to  temporal correlation, the quality information  computed and incorporated into ADMM iterations further improves the performance of ADMM based MUD for the proposed dynamic system model.



 \section{Codebook based Grant free SIMO CD-NOMA system model} \label{sec_II}
 In this section, we discuss the block-sparsity models proposed for grant-free SCMA and DCMA systems. The inherent block-sparsity structure in grant-free codebook-based CD-NOMA systems leverages BCS to enhance the performance of JADD. This underlying block-sparsity structure in SCMA and DCMA arises from the codewords used during the encoding process. Each codeword contains a cluster of non-zero elements that represent the information bits of a single user. The clustered nature of codebook-based CD-NOMA systems motivates the design of block-sparsity-based SCMA and DCMA models. Subsection~\ref{sub_A} introduces a one-shot block-sparsity model that considers a single time slot for transmission. Subsection~\ref{sub_c} presents a dynamic frame-wise model, where users access the network randomly over time.
 \subsection{One shot static block sparsity model}\label{sub_A}
In this model, we consider an uplink (UL) grant-free codebook-based CD-NOMA system with a single time slot for transmission. The users transmitting data in this time slot are considered active, while the remaining users are inactive. Active users transmit a codeword, while inactive users remain silent and are assumed to transmit a block of zeros during the time slot. Consequently, the augmented joint codebook for all users, denoted as $\mathcal{X}\triangleq{\mathcal{X}_a\cup 0}$, represents the codewords for both active and inactive users.
 The observation on $n_{\rm r}$th antenna at the BS
 \begin{align*}
 	\mathbf{r}^{(n_{\rm r})}=& \sum_{j=1}^{J} \text{diag}(\mathbf{h}_j^{(n_{\rm r})})\mathbf{x}_j+\mathbf{w^{(n_{\rm r})}}			
 \end{align*}
 The entire observation at BS is given by

 \begin{align}\label{eq_GF}
 	[\mathbf{r}]_{N_{\rm r}K\times 1}=\mathbf{H}\mathbf{x}+\mathbf{w}
 \end{align}
 where
 \begin{itemize}
 	\item $\mathbf{r}=\left[ \mathbf{r}^{{(1)}^T},\mathbf{r}^{{(2)}^T},\hdots,\mathbf{r}^{{(N_{\rm r})}^T}   \right]$
 	\item $\mathbf{w^{(n_{\rm r})}}\sim \mathcal{CN}(0,\sigma^2\textbf{I}_{KN_{\rm r}})$ denotes the additive white Gaussian noise (AWGN) at the receiver.
 	\item $\mathbf{h}_j^{n_{\rm r}}\sim \mathcal{CN}(0,\textbf{I}_K)$ denotes the Rayleigh fading  channel vector for the $j$th UE.
 	\item  		\begin{align*}\textbf{H}=
 		\begin{bmatrix}
 			\text{diag}(\textbf{h}_1^{(1)}),\hdots,\text{diag}(\textbf{h}_j^{(1)}),\hdots,\text{diag}(\textbf{h}_J^{(1)})\\
 			\text{diag}(\textbf{h}_1^{(2)}),\hdots,\text{diag}(\textbf{h}_j^{(2)}),\hdots,\text{diag}(\textbf{h}_J^{(2)})	\\
 			\\
 			\\
 			\text{diag}(\textbf{h}_1^{(N_r)}),\hdots,\text{diag}(\textbf{h}_j^{(N_r)}),\hdots,\text{diag}(\textbf{h}_J^{(N_r)})									
 		\end{bmatrix}   
 	\end{align*}
 	\item $ \mathbf{x}=\left[\mathbf{x}_1^T,\mathbf{x}_2^T,\hdots,\mathbf{x}_J^T \right]_{JK\times 1}^T$, is a joint user codeword vector  and  $\mathbf{x}_j=\left[x_{j,1},x_{j,2},\hdots,x_{j,K}\right]_{K\times 1}^T$.
 \end{itemize}
The transmitted signal $\mathbf{x}$ in a codebook-based grant-free CD-NOMA system inherently exhibits a block-sparsity structure, with each user's codeword serving as a distinct block corresponding to that user. In the context of JADD, user activity detection effectively leverages the block-wise sparsity provided by the CD-NOMA codewords.
The active user set is defined as 
$$\Lambda=\big\{ j:j\in \{1,2,\hdots,J\}\mid\Vert\mathbf{x}_j\Vert_2\neq 0\big\}$$
The sparsity level denoted as  $\rm S_{\rm l}$, is defined as $\rm S_{\rm l}=|\Lambda|$.

\subsection{Frame-wise static block sparsity model}\label{sub_b}
Frame-wise sparsity structure improves the performance of MUD over on-shot static model. 
Frame-wise wise static block sparsity model is an extension of the one-shot static block sparsity model to multiple time slots. Thus, the data is transmitted from the active user in a continuous fashion. In turn, the user is active or inactive throughout the frame. This model doesn't considered the random activity of the users within the frame. The active users in each time slot of frame remains unchanged. The frame-wise sparsity structure is standard to model the continuous data transfer in spreading sequence based CD-NOMA. However, the block-sparsity structure of grant-free codebook based CD-NOMA leads to  frame-wise block sparsity model.  The frame-wise sparsity enjoys the performance gains over the one shot static model discussed in Section \ref{s_3}.
\begin{align}
	[\mathbf{r}]_{N_{\rm r}K\times 1}^{[l]}=\mathbf{H}^{[l]}\mathbf{x}^{[l]}+\mathbf{w}^{[l]},\quad l=1,\hdots,L.\label{eq_Dy}
\end{align}
where $L$ is the frame-size.
The frame-wise static block-sparsity model can be expressed as
\begin{align}
\mathbf{R}=\mathbf{H}\mathbf{X}+\mathbf{W}
\end{align} 
where $\mathbf{R}=[\mathbf{r}^{[1]},\mathbf{r}^{[2]},\hdots,\mathbf{r}^{[L]}]$, $\mathbf{H}^{[l]}=\mathbf{H},\forall l$, indicates the slow fading channel, 
$\mathbf{X}=[\mathbf{x}^{[1]},\mathbf{x}^{[2]},\hdots,\mathbf{x}^{[L]}]$. The input matrix $\mathbb{X}$ maintains the frame-wise sparse structure as shown in Figure.

The frame-wise static  model is given as
$$\Lambda^{[1]}=\Lambda^{[2]}=\hdots=\Lambda^{[L]}$$   
where $\Lambda^{[l]}=\big\{ j:j\in \{1,2,\hdots,J\}\mid\Vert\mathbf{x}_j^{[l]}\Vert\neq 0\big\}$ indicates the set of active users in $l$th time-slot. The sparsity level is defined as the  cardinality of the set $\Lambda^{[l]}$ and which remains unchanged throughout the frame. 

The joint processing of the frame of observations $\mathbf{r}^{[l]}, l=1,\hdots, L$ helps in improving the accuracy of the activity detection. 
Thus, the frame-wise sparsity enables joint sparse recovery of $\mathbf{x}^{[l]}, l=1,\hdots, L$ to improve the performance of MUD over one-shot static model.

\subsection{Frame-wise dynamic block-sparsity model}\label{sub_c}

In this subsection, we present a practical system model that is adaptable to massive machine-type communications (mMTCs). The one-shot static model ignores the critical aspect: the temporal correlation present in grant-free NOMA systems. A frame-wise sparsity structure is not appropriate when user activity is dynamic over time. Therefore, we need to design a system that accommodates varying user activity across frames.

The title of this section integrates three concepts—block sparsity, dynamism, and frame-wise design—to develop a system model for practical grant-free codebook-based CD-NOMA systems. The block-sparsity model introduced here is inspired by the one-shot static model discussed in Subsection~\ref{sub_A}. This model incorporates a frame-wise dynamic transmission approach within the uplink codeword-based CD-NOMA system, capturing the dynamism present in sporadic traffic networks, such as mMTCs. In real-world scenarios, the set of active users in the network changes continuously and gradually. A frame consists of multiple time slots, allowing users to enter or leave the network at any time. This temporal correlation can be leveraged to detect active users using the dynamic block-compressive sensing (BCS) algorithm discussed in the following section.  The entire observation at BS in time slot $t$ is given by

\begin{align}
	[\mathbf{r}]_{N_{\rm r}K\times 1}^{[l]}=\mathbf{H}^{[l]}\mathbf{x}^{[l]}+\mathbf{w}^{[l]},\quad l=1,\hdots,L.\label{eq_Dy}
\end{align}
where $\mathbf{x}^{[l]}$  may vary across different time slots and the active user support set in (\ref{eq_Dy}) is given as
$$\Lambda^{[1]}\neq\Lambda^{[2]}\neq\hdots=\Lambda^{[L]}$$   

The system model~(\ref{eq_Dy} ) indicates the observation in each time slot is different. Thus, the joint processing of the frame of observations is not possible. The objective of JADD problem for (\ref{eq_Dy}) is to detect the active users and data in each time slot. The observation in each time slot $\mathbf{r}^{[l]}$ help recover the $\mathbf{x}^{[l]}$ in the corresponding time slot. However, the temporal correlation of active user sets in continuous time slots can be exploited for the active user set detection. Most of the users in the previous time slots might be common in the present time slot due to the continuous data transmission. The estimated active user set and the data  in the previous time slot can be acts as the prior information for the present time slot.  The BCS based JADD that incorporates the prior information improves the performance of frame-wise dynamic block-sparsity system model.

The dynamic model considered in this paper discuss two different use cases exists in the mMTC network.
\begin{enumerate}
	\item Active user activity changes at a very fast rate.
	\item Active user activity  changes at a slow rate.
\end{enumerate}
\textbf{ Case-1: }In the first case, usually occurs in short packet transmissions, where the user is active for very short span of time. Thus, it leads to the assumption in which the use activity time strictly less than the duration of the time slot in a frame. Hence, the user activity among several time slots is independent. The JADD in each time slot is to be performed independently. 

 \textbf{ Case-2: }
In the second case, this typically occurs during moderate packet transmissions or when the time slot within a frame is of short duration. As a result, the user's activity period exceeds the length of the time slot. Some active users, with high probability, may transmit data across adjacent time slots. This creates a temporal correlation between the sets of active users in consecutive time slots, which can be leveraged to enhance MUD (multi-user detection) performance in two ways:
\begin{itemize}
	\item \textbf{Case 2a:} When the symbol duration is shorter than the time slot (i.e., symbol duration < time slot), temporal correlation exists between time slots only through the active user sets.
	\item\textbf{ Case 2b:} When the symbol duration exceeds the time slot (i.e., symbol duration > time slot), temporal correlation is present not only in the active user sets but also in the signal information.
\end{itemize}
 The detailed formulation and algorithm of JADD for (\ref{eq_GF}) and (\ref{eq_Dy}) discussed in Section~\ref{s_3}.
 \section{ADMM based joint activity and data detection for (\ref{eq_GF}) and (\ref{eq_Dy})}\label{s_3}
 The system model for SIMO CD-NOMA is given in (\ref{eq_GF}). Wherein the transmit signal vector ($\mathbf{x}$) is concatenation of $J$-user codewords. Assume a mMTC scenario, where only few users are active and all other users don't involve in the transmission process. Further, base station (Bs) doesn't have any information about the set of active users. The challenge is to detect the  set of active users correctly followed by data detection known as JADD. The codebook based CD-NOMA system holds the block sparsity structure in transmit vector~\cite{Elder_block}. The block sparsity is due to the non-zero entries in the codewords occurring in clusters. Further, each codeword or block corresponding to one user. There are two different types of variants available in codebook based CD-NOMA system. One is sparse CD-NOMA system, which utilizes the sparse codebooks and another one  is dense CD-NOMA system, which utilizes the dense codebooks in the CD-NOMA encoding. The $j$th user codeword $\mathbf{x}_j$ is sparse for sparse CD-NOMA system and dense for dense CD-NOMA system.
 The sparsity within the codeword or block in the sparse system need to be treated separately, unlike dense CD-NOMA system. It makes the difference in the JADD problem formulation as discussed in this section. Throughout this work the sparse CD-NOMA systems refereed as SCMA and dense CD-NOMA systems referred as DCMA.  This section presents ADMM based JADD for both SCMA and DCMA systems.
\subsection{JADD for one-shot block sparsity model(\ref{eq_GF})}\label{sub_SGLASSO}
The focus of this subsection is to design a multi-user detector (MUD) for JADD for the one-shot static block sparsity system discussed in subsection~\ref{sub_A}. This section includes the detection problem formulation for both sparse as well as dense CD-NOMA systems.
 \subsubsection{JADD for sparse CD-NOMA system}\label{JADD_SG_LASSO}
 SCMA is one of the efficient codebook based sparse CD-NOMA system. This section discusses the multi-user detection  problem formulation for SCMA systems in grant-free scenario.
 As mentioned earlier, grant free NOMA systems allows the user access without any Grant from the base station (BS). It leads to multiple users to occupy the same resource. Further,  BS doesn't have any information about the active users at any given point of time in the network. Consider a CD-NOMA system having total $J$ users in the system, in the only $J_a$ are the active users, where $J_a<<J$ and remaining all users are in silent mode. The SCMA system leads to user-wise sparsity and within user sparsity. The user-wise sparsity is due to many users in the system are not active, no data has been transmitted from those users. The within user sparsity is due to the active users transmit the sparse codewords with only few non-zero entries. The detection problem of grant-free SCMA system can be formulated as sparse group LASSO problem. The ADMM algorithm is proposed to solve the formulated sparse group LASSO problem in a distributed manner.
 Sparse group LASSO for SCMA system according to (\ref{eq_GF})
 \begin{align}
 	\min_{\mathbf{x}}\quad \Vert \mathbf{r}-\mathbf{H}\mathbf{x}\Vert^2+\underbrace{\alpha_1\sum_{j=1}^J\Vert\mathbf{x}_j\Vert_2}_{\text{user wise sparsity}}+\underbrace{\alpha_2\Vert\mathbf{x}\Vert_1}_{\text{within  user sparsity }}\label{eq_LASS1}
 \end{align}
The problem (\ref{eq_LASS1})  includes three terms. The  first term indicates the data fidelity term and remaining two terms indicate the penalty terms which promotes the sparse structure of $\mathbf{x}$. The first penalty is the mixed $\ell_{2,1}$ norm, which promotes the user wise sparsity to refer to the active users in the system. The second penalty is the $\ell_1$ norm, which promotes the within user sparsity or overall sparsity to refer to the sparse codewords in SCMA system. These two penalty terms helps in finding the solution with fewer non-zero entries along with the block sparsity structure. 
 The $\alpha_1>0$ and $\alpha_2>0$ represents the regularization or penalty parameters for activity and data detection, respectively. It helps to tune the sparsity of the solution. The problem (\ref{eq_LASS1}) indicates that the objetive function is convex with two non-smooth penalty terms. It can be solved by using the sub-differential calculus as discussed in the ADMM algorithm.

  The optimum selection of the penalty parameters improves the performance of the detector. The problem in (\ref{eq_LASS1}) ignoring the fact that the imposition of penalty on active users must be lesser than the active users. The uniform penalization used in (\ref{eq_LASS1}) leads to losses in the performance. Hence, a non-uniform or reweighed strategy of penalization is adapted in this work.
  \begin{align}
 	\min_{\mathbf{x}}\quad \Vert \mathbf{r}-\mathbf{H}\mathbf{x}\Vert^2+\underbrace{\alpha_1\sum_{j=1}^Jw_j\Vert\mathbf{x}_j\Vert_2}_{\text{user wise sparsity}}+\underbrace{\alpha_2\Vert\mathbf{x}\Vert_{1,w}}_{\text{within  user sparsity }}\label{eq_LASS}
 \end{align}
 where $\Vert\mathbf{x}\Vert_{1,w}=\sum_{i=1}^{JK} w_i\vert x_i\vert$, $x_i$ is the $i$th entry of the vector $\mathbf{x}$. The weights $w_j$ and $w_i$ are selected as inversely propositional to the  norm and absolute values of the $\mathbf{x}_j$ and $x_i$ of the previous iteration as follows
 \begin{align}
 	w_j^t=\frac{1	}{\Vert \mathbf{x}_j^{t-1}\Vert+\epsilon}\quad w_i^t=\frac{1}{\vert x_i^{t-1}\vert+\epsilon}.\label{rew}
 \end{align}
According to (\ref{rew}), the users and the codeword entries  which are having low magnitude values in the previous iteration penalize more in the present iteration.  When the iteration progresses the penalty imposing on the  active users (first penalty) and noz-zero entries (second penalty) approaching to zero. \\
\textbf{ADMM algorithm:}\\
 The ADMM for the problem in (\ref{eq_LASS}) easy to implement and guaranteed convergence property~\cite{boyd}. For the ADMM formulation, split the variable $\mathbf{x}$ by introducing two auxiliary variables $\mathbf{q}$ and $\mathbf{z}$.
 The ADMM problem formulation is given by
 \begin{align}
 	\min_{\mathbf{x},\mathbf{z},\mathbf{q}}\quad &\Vert \mathbf{r}-\mathbf{H}\mathbf{x}\Vert^2+\alpha_1\sum_{j=1}^Jw_j\Vert\mathbf{z}_j\Vert_2 +\alpha_2\Vert\mathbf{q}\Vert_{1,w}\label{eq_LASSO}\\
 	\text{s.t} \quad&\mathbf{x}=\mathbf{z}, \mathbf{x}=\mathbf{q}\nonumber
 \end{align}

 The augmented Lagrangian function for (\ref{eq_LASSO}) is given by
 
 \begin{align}\label{eq_aug}
 	\begin{split}
 	\mathcal{L}\left(\mathbf{x},\{\mathbf{z}_j\}_{j=1}^J,\mathbf{q},\mathbf{y}_1,\mathbf{y}_2\right)=	\frac{1}{2}\Vert \mathbf{r}-\mathbf{H}\mathbf{x}\Vert^2+\alpha_1\sum_{j=1}^Jw_j\Vert\mathbf{z}_j\Vert_2\\ +\alpha_2\Vert\mathbf{q}\Vert_{1,w}+ {\text{Re}}\innerproduct{\mathbf{y}_1}{\mathbf{x}-\mathbf{z}}+
 	{\text{Re}}\innerproduct{\mathbf{y}_2}{\mathbf{x}-\mathbf{q}}\\+\frac{\rho}{2}\Vert \mathbf{x}-\mathbf{z}\Vert^2+\frac{\rho}{2}\Vert \mathbf{x}-\mathbf{q}\Vert^2
	 \end{split}
 \end{align}


 where $\mathbf{x},\mathbf{q}$, and $\mathbf{z}$ are the primal variables and $\mathbf{z}=~[\mathbf{z}_1^T,\mathbf{z}_2^T,\hdots,\mathbf{z}_J^T]$ , $\mathbf{y}_1$ and $\mathbf{y}_2$ are the Lagrangian multipliers and $\rho>0$ is the augmented Lagrangian penalty parameter. 
 \begin{subequations}
 \begin{align}\small
 	\begin{split}
 		\mathbf{x}^{t+1}:={}&\argmin_{\mathbf{x}}\mathcal{L}\left(\mathbf{x},\{\mathbf{z}_j^t\}_{j=1}^J,\mathbf{q^t},\mathbf{y}_1^t,\mathbf{y}_2^t\right)\\
 		:=&\argmin_{\mathbf{x}} \frac{1}{2}\Vert \mathbf{r}-\mathbf{H}\mathbf{x}\Vert^2+{\text{Re}}\innerproduct{\mathbf{y}_1}{\mathbf{x}-\mathbf{z}}\\&+
 		{\text{Re}}\innerproduct{\mathbf{y}_2}{\mathbf{x}-\mathbf{q}}+\frac{\rho}{2}\Vert \mathbf{x}-\mathbf{z}\Vert^2+\frac{\rho}{2}\Vert \mathbf{x}-\mathbf{q}\Vert^2
 	\end{split}\label{eq_7a}\\
 	\begin{split}
 		\mathbf{q}^{t+1}:={}&\argmin_{\mathbf{q}}\mathcal{L}\left(\mathbf{x}^{t+1},\{\mathbf{z}_j^t\}_{j=1}^J,\mathbf{q},\mathbf{y}_1^t,\mathbf{y}_2^t\right)\\
 		:=&\argmin_{\mathbf{q}} \alpha_2\Vert\mathbf{q}\Vert_{1,w}+{\text{Re}}\innerproduct{\mathbf{y}_2}{\mathbf{x}-\mathbf{q}}+\frac{\rho}{2}\Vert \mathbf{x}-\mathbf{q}\Vert^2
 	\end{split}\label{eq_7b}\\
 	\mathbf{z}^{t+1}:={}& \mathcal{L}\left(\mathbf{x}^{t+1},\{\mathbf{z}_j\}_{j=1}^J,\mathbf{q}^{t+1},\mathbf{y}_1^t,\mathbf{y}_2^t\right)\\
 	:=&\argmin_{\mathbf{z}} \alpha_1\sum_{j=1}^Jw_j\Vert\mathbf{z}_j\Vert_2+{\text{Re}}\innerproduct{\mathbf{y}_1}{\mathbf{x}-\mathbf{z}}\nonumber\\&+\frac{\rho}{2}\Vert \mathbf{x}-\mathbf{z}\Vert^2 \label{eq_7c}\\
 	\mathbf{y}_1^{t+1}:={}&\mathbf{y}_1^{t}+\rho(\mathbf{z}^{t+1}-\mathbf{x}^{t+1})\\
 \mathbf{y}_2^{t+1}:={}&\mathbf{y}_2^{t}+\rho(\mathbf{q}^{t+1}-\mathbf{x}^{t+1})
 \end{align}
\end{subequations}
 
 Letting $\mathbf{u}_1=\frac{\mathbf{y}_1}{\rho}$ and $\mathbf{u}_2=\frac{\mathbf{y}_2}{\rho}$, called the dual variables~\cite{boyd}. The scaled version of ADMM steps as follows
 \begin{subequations}\label{eq8_set}
 \begin{align}\small
 	\begin{split}
 		\mathbf{x}^{t+1}:={}&\argmin_{\mathbf{x}} \frac{1}{2}\Vert \mathbf{r}-\mathbf{H}\mathbf{x}\Vert^2+\frac{\rho}{2}\Vert \mathbf{x}-\mathbf{z}+\mathbf{u}_1\Vert^2\\&+\frac{\rho}{2}\Vert \mathbf{x}-\mathbf{q}+\mathbf{u}_2\Vert^2\label{eq_9}
 	\end{split}\\
 	\begin{split}
 		\mathbf{q}^{t+1}:={}&\argmin_{\mathbf{q}} \alpha_2\Vert\mathbf{q}\Vert_{1,w}+\frac{\rho}{2}\Vert \mathbf{x}-\mathbf{q}+\mathbf{u}_2\Vert^2\label{eq_10}
 	\end{split}\\
 	\mathbf{z}^{t+1}:={}&\argmin_{\mathbf{z}} \alpha_1 \sum_{j=1}^J w_j\Vert\mathbf{z}_j\Vert_2+\frac{\rho}{2}\Vert \mathbf{x}-\mathbf{z}+\mathbf{u}_1\Vert^2\label{eq_11}\\ 		
 	\mathbf{u}_1^{t+1}:={}&\mathbf{u}_1^{t}+(\mathbf{z}^{t+1}-\mathbf{x}^{t+1})\\
 	\mathbf{u}_2^{t+1}:={}&\mathbf{u}_2^{t}+(\mathbf{q}^{t+1}-\mathbf{x}^{t+1})
 \end{align}
\end{subequations}
 The involved augmented Lagrangian function in  (\ref{eq_aug}) is now decomposed into simple optimization problems (\ref{eq8_set}). The first three steps corresponds to primal variable optimization and the remaining two steps corresponds to  dual variable update.
 The problem  (\ref{eq_11}) can be simplified and solved distributively for $j=1,\hdots,J$, the equation (\ref{eq_11}) can be equivalently written as
 \begin{align}
 	\sum_{j=1}^J \min_{\mathbf{z}_j} \alpha_1 w_j \Vert\mathbf{z}_j\Vert_2+\frac{\rho}{2}\Vert\mathbf{x}_j-\mathbf{z}_j+\mathbf{u}_{1j}\Vert^2\label{eq_141}
 \end{align}
 The problem (\ref{eq_141}) can be separately solved for each $j$ w.r.t the variable $\mathbf{z}_j$.  The main challenge is to compute the optimum solutions for the optimization problems in (\ref{eq8_set}). For the problem (\ref{eq_9}) simplified from (\ref{eq_7a}), it can be observed that $\mathcal{L}\left(\mathbf{x},\{\mathbf{z}_j^t\}_{j=1}^J,\mathbf{q^t},\mathbf{y}_1^t,\mathbf{y}_2^t\right)$ is convex w.r.t $\mathbf{x}$ when $\rho>0$ and $\mathbf{H}^H\mathbf{H}$ is positive semi definite (PSD). Further, it is strongly convex w.r.t $\mathbf{x}$ when it holds $\mathbf{H}^H\mathbf{H}+\rho I\succeq(\lambda_{min}(\mathbf{H}^H\mathbf{H})+\rho)I$.
 For a problem (\ref{eq_10}) derived from (\ref{eq_7b}), $\mathcal{L}\left(\mathbf{x}^{t+1},\{\mathbf{z}_j^t\}_{j=1}^J,\mathbf{q},\mathbf{y}_1^t,\mathbf{y}_2^t\right)$ is  convex w.r.t the variable $\mathbf{q}$ when $\rho>0$ and eventually strongly convex as $\rho$  increases  . For a problem (\ref{eq_11}) derived from (\ref{eq_7c}), $\mathcal{L}\left(\mathbf{x}^{t+1},\{\mathbf{z}_j\}_{j=1}^J,\mathbf{q}^{t+1},\mathbf{y}_1^t,\mathbf{y}_2^t\right)$
  is convex w.r.t the variable $\mathbf{z}$ when $\rho>0$ and eventually strongly convex  due to its second term when $\rho>0$. Therefore, optimum solution for each of the above optimization problems can be obtained by setting the gradient w.r.t corresponding variables equal to zero as following:
  \begin{align}
  	\nabla_{\mathbf{x}} \mathcal{L}\left(\mathbf{x},\{\mathbf{z}_j^t\}_{j=1}^J,\mathbf{q^t},\mathbf{y}_1^t,\mathbf{y}_2^t\right)=0,\\
  	\partial_{\mathbf{q}}\mathcal{L}\left(\mathbf{x}^{t+1},\{\mathbf{z}_j^t\}_{j=1}^J,\mathbf{q},\mathbf{y}_1^t,\mathbf{y}_2^t\right)\in 0,\\
  	\partial_{\mathbf{z}_j}\mathcal{L}\left(\mathbf{x}^{t+1},\{\mathbf{z}_j\}_{j=1}^J,\mathbf{q}^{t+1},\mathbf{y}_1^t,\mathbf{y}_2^t\right)\in 0.  	
  	\end{align}
  where $\partial$ indicates the subgradient operation.
 The first term in (\ref{eq_10}) and   (\ref{eq_11}) are non-differentiable, that can be solved with sub-differential methods. The solution to (\ref{eq_10}) and (\ref{eq_11}) by using scalar soft thresholding operation and block soft thresholding operation  popularly known as shrinkage operation.

 The steps to solve the problem (\ref{eq_141}) is provided in Appendix.
 The ADMM solutions for the problems (\ref{eq_9}), (\ref{eq_10}), and (\ref{eq_11}) as follows 
 \begin{subequations}
 \begin{align}	\small	
 	\mathbf{x}^{t+1}:={}& (\mathbf{H}^{\rm H}\mathbf{H}+2\rho \mathbf{I})^{-1}(\mathbf{H}^{H}\mathbf{y}+\rho(\mathbf{z}^k+\mathbf{q}^k-\mathbf{u}_1^k-\mathbf{u}_2^k))\label{eq_13a}\\
 	\mathbf{q}^{t+1}:={}&\text{max}(	\mathbf{x}^{t+1}+\mathbf{u}_2^k-\frac{\alpha_2 w_i^t}{\rho},0)\nonumber\\&-\text{max}(-\mathbf{x}^{t+1}-\mathbf{u}_2^k-\frac{\alpha_2w_i^t}{\rho},0)\label{eq_13b}\\
 	\mathbf{z}_j^{t+1}:={}&\frac{(\mathbf{x}_j^{t+1}+\mathbf{u}_{1j}^{t})}{\Vert \mathbf{x}_j^{t+1}+\mathbf{u}_{1j}^{t}\Vert}\bigg( \Vert \mathbf{x}_j^{t+1}+\mathbf{u}_{1j}^{t}\Vert-\frac{\alpha_1w_j^t}{\rho} \bigg)_+,\quad\forall j\label{eq_13c}\\
 	\mathbf{u}_1^{t+1}:={}&\mathbf{u}_1^t+(\mathbf{x}^{t+1}-\mathbf{z}^{t+1})\label{eq_13d}\\
 	\mathbf{u}_2^{t+1}:={}&\mathbf{u}_2^t+(\mathbf{x}^{t+1}-\mathbf{q}^{t+1})\label{eq_13e}
 \end{align}
\end{subequations}

\begin{algorithm}[t]
	\small
	\caption{Proposed group LASSO based JADD via FSJ aided ADMM algorithm for Model-1}
	\begin{algorithmic}[1]
		\State  \textbf{Input:} $\mathbf{r}$,$\mathbf{H}$, and maximum number of iterations ($T$). \\	
		\textbf{Output:} $ \hat{\Lambda}$, and $\hat{\mathbf{x}}$.
		\State \textbf{Initialization:} Initialize the vectors $\mathbf{z}$,$\mathbf{q}$,and $\mathbf{u}_1$ with zeros.
		\State \textbf{Pre-processing}\\
		Compute $(\mathbf{H}^H\mathbf{H}+2\rho\mathbf{I})^{-1}$.\\
		Compute $\mathbf{H}^H\mathbf{r}$. 		
		\For {$t=1,2,,\hdots T $}
		\State \textbf{Step}:1 Update $ \mathbf{x}^{(t+1)}$ via (\ref{eq_22a}),
		\State  \textbf{Step}:2
		Update $\mathbf{z}_j^{(t+1)}$ via (\ref{eq_22b}), for $j=1,2,\hdots,J$
		
		\State \textbf{Step}:3 Update $\mathbf{u}_1^{(t+1)}$ via (\ref{eq_22c}),\\
		(Computing primal residuals)
		\begin{align*}
			\mathbf{r}_{p}^{(t+1)}=\mathbf{x}^{(t+1)}-\mathbf{z}^{(t+1)},\quad \Vert \mathbf{r}_{p}^{(t+1)} \Vert_2
		\end{align*}
		(Computing dual residual)
		\begin{align*}
			\mathbf{r}_d^{(t+1)}=\rho(\mathbf{z}^{(t)}-\mathbf{z}^{(t+1)})\quad \Vert	\mathbf{r}_d^{t+1}\Vert_2	
		\end{align*}	
	$$\epsilon^{\rm pri}=\sqrt{J*K}\epsilon^{\rm abs}+\epsilon^{\rm rel} \max\lbrace \Vert\mathbf{x^{(t+1)}}\Vert_2,\Vert \mathbf{z^{(t+1)}}\Vert_2\rbrace$$
	$$\epsilon^{\rm dual }=\sqrt{J*K}\epsilon^{\rm abs}+\epsilon^{\rm rel} \Vert\rho \mathbf{u}_1^{(t+1)}\Vert $$
		\If{$\Vert \mathbf{r}_{p}^{(t+1)} \Vert_2\leq \epsilon^{\rm pri} \land\mathbf{r}_d^{(t+1)}\leq  \epsilon^{\rm dual }  $}
		\\\quad Stop the iteration
		\EndIf
	
		\EndFor
		
		\State \textbf{Post-iteration processing}\\
		(Computing FSJ)\\
		$\widetilde{\mathbf{x}}=[\Vert \mathbf{x}_1^{(T)}\Vert_2,\Vert \mathbf{x}_2^{(T)}\Vert_2,\hdots,\Vert \mathbf{x}_J^{(T)}\Vert_2]$\\
		First sort the elements of $\widetilde{\mathbf{x}}$ in ascending order,
		$$\widetilde{\mathbf{x}}_{\rm sort}=\rm sort(\widetilde{\mathbf{x}})$$
		i.e. $\widetilde{\mathbf{x}}_{\rm sort}[1]\leq\widetilde{\mathbf{x}}_{\rm sort}[2]\leq,\hdots,\leq,\widetilde{\mathbf{x}}_{\rm sort}[J]$, where $\widetilde{\mathbf{x}}_{\rm sort}[p]$ indicates the $p$-th largest entry in $\widetilde{\mathbf{x}}_{\rm sort}$.\\
		The FSJ computed from the consecutive difference of sorted $\hat{\mathbf{x}}_{\rm sum}$ is given by
		\begin{align}
			\widetilde{\mathbf{x}}_{\rm sort}[p+1]-\widetilde{\mathbf{x}}_{\rm sort}[p]>\frac{\alpha\Vert\widetilde{\mathbf{x}} \Vert_{\infty}}{N_{\rm r}}\label{FSJ11}
		\end{align}
		The smallest $p$ in (\ref{FSJ11}) indicates the FSJ, then $\beta=~\widetilde{\mathbf{x}}_{\rm sort}[p]$. and the partial support set 
		\begin{align}
			\mathcal{T}=\lbrace j: \widetilde{\mathbf{x}}_j>\beta \rbrace\label{sup}
		\end{align}
	  Sparsity level $\hat{\rm S}_{\rm l}=\vert \mathcal{T} \vert$.
		\State  (Computation of final support set)\\
		$\widetilde{\mathbf{x}}=[\Vert \mathbf{x}_1\Vert_2,\Vert \mathbf{x}_2\Vert_2,\hdots,\Vert \mathbf{x}_J\Vert_2]$\\
		Find out the set of $\hat{\rm S}_{\rm l}$ user indices in $\widetilde{\mathbf{x}}$, whose $\Vert \mathbf{x}_j\Vert_2$ values are largest among $j=1,2,\hdots,J$ is indicated by
		$\Xi(\widetilde{\mathbf{x}},\hat{\rm S}_{\rm l})$
		(Final detected support set)\\		
		$\hat{\Lambda}=\Xi(\widetilde{\mathbf{x}},\hat{\rm S}_{\rm l})$.\\
		(Final estimated signal information)\\
		$\hat{\mathbf{x}}=\mathbf{x}^{(T)}[\lbrace 1,2,\hdots,J  \rbrace \setminus\hat{\Lambda}]=0$.

	\end{algorithmic} 
	\label{alg1}
\end{algorithm}
A threshold-aided support detection scheme for the sparse recovery problem is proposed in \cite{FSJ}, based on the principle of the first significant jump (FSJ). This concept is applied in Algorithm~\ref{alg1} for support detection when the true sparsity level is unknown. Subsequently, it is adapted for Algorithm~\ref{alg2}, which is designed for dynamic scenarios. The FSJ occurs within the estimated sequence, where the true non-zero elements are large in magnitude and few in number, while the false ones are small in magnitude and more numerous. This creates a clear boundary between the true non-zero and false non-zero elements.

\begin{algorithm}[t]
	\small
	\caption{JADD for frame-wise dynamic block-sparsity model via FSJ-ADMM for Model-2}
	\begin{algorithmic}[2]
		\State  \textbf{Input:} $\mathbf{r}^{[l]}$,$\mathbf{H}^{[l]}$  for $l=1,2,\hdots,L.$ \\	
		\textbf{Output:} $ \hat{\Lambda}^{[l]}$, and $\hat{\mathbf{x}}^{[l]}$ for $l=1,2,\hdots,L$.
		\State \textbf{Initialization:} Initialize the vectors $\mathbf{z}$,$\mathbf{q}$,$\mathbf{u}_1$, and $\mathbf{u}_2$ with zeros.
		
		\State \textbf{Step-i:} Repeat \textbf{Algorithm}~\ref{alg1} for $l=1,2,\hdots,L$ independently and compute $\hat{\Lambda}_{\rm p}^{[l]}$ and $\hat{\mathbf{x}}^{[l]},\forall l$.
		$$\hat{\mathbf{X}}=[\hat{\mathbf{x}}^{[1]},\hat{\mathbf{x}}^{[2]},\hdots, \hat{\mathbf{x}}^{[L]}]$$
		
		\State Firstly perform operation in (\ref{step-2})
		
		\If{$\hat{\mathbf{X}}\in\text{Case-1}$ in (\ref{step-2})}\\
		Stop the algorithm
		\Else\\
		Continue to \textbf{Step-ii}
		
				\If{$\hat{\mathbf{X}}\in\text{Case-2a}$ in (\ref{step-2_case2b})}\\
				Apply Algorithm~\ref{alg1} to the problem given in (\ref{eq_step2}) for $l=1,2,\hdots,L$\\
				(Final detected support set)\\		
				$\hat{\Lambda}_{\rm fin}^{[l]}=\Xi(\widetilde{\mathbf{x}},\hat{\rm S}_{\rm l}^{[l]})$.\\
				
					(Final estimated signal information)\\
					$\hat{\mathbf{x}}_{\rm fin}^{[l]}=\mathbf{x}^{(T)}[\lbrace 1,2,\hdots,J  \rbrace \setminus\hat{\Lambda}_{\rm fin}^{[l]}]=0$, for $l=1,2,\hdots,L$.
				
				\ElsIf  {$\hat{\mathbf{X}}\in\text{Case-2b}$ in (\ref{step-2_case2b})}\\
				 
				 Apply Algorithm~\ref{alg1} to the problem given in (\ref{eq_Case2b}) by incorporating the corresponding solutions given in (\ref{case_2b_steps})  for $l=1,2,\hdots,L.$\\	
				 (Final detected support set)\\		
				 $\hat{\Lambda}_{\rm fin}^{[l]}=\Xi(\widetilde{\mathbf{x}},\hat{\rm S}_{\rm l}^{[l]})$.\\
				 
				 	(Final estimated signal information)\\			 
				 $\hat{\mathbf{x}}_{\rm fin}^{[l]}=\mathbf{x}^{(T)}[\lbrace 1,2,\hdots,J  \rbrace \setminus\hat{\Lambda}^{[l]}]=0$, for $l=1,2,\hdots,L$.
				
				\EndIf

		\EndIf

	\end{algorithmic} 
	\label{alg2}
\end{algorithm}

 \subsubsection{JADD for dense CD-NOMA system}
 This subsection focuses on the detection formulation for dense CD-NOMA systems popularly known as DCMA systems. The JADD problem for DCMA system is modified by keeping $\alpha_2=0$, means there is no within user sparsity exists in the DCMA.  Thus, the DCMA codewords are dense and clustered non-zero values occurs in dense manner within the  codeword. 
 \begin{align}
 	\min_{\mathbf{x}}\quad \Vert \mathbf{r}-\mathbf{H}\mathbf{x}\Vert^2+\underbrace{\alpha_1\sum_{j=1}^J w_j\Vert\mathbf{x}_j\Vert_2}_{\text{user wise sparsity}}\label{eq_LASS_DCMA}
 \end{align}
  The problem (\ref{eq_LASS_DCMA})  similar  to the problem in (\ref{eq_LASS}) by keeping $\alpha_2=0$. Sparse group LASSO problem is now converted into group LASSO problem.  Group LASSO problem helps in computing the solution with sparse set of blocks, if it includes a user (active) in the transmission then all the entries of that block will be non zero. By following the steps (\ref{eq_LASSO}) and (\ref{eq_aug}), the 3-ADMM steps are derived as follows
   \begin{subequations}
   \begin{align}\small
  	\begin{split}\label{eq_18a}
  		\mathbf{x}^{t+1}:={}&\argmin_{\mathbf{x}} \frac{1}{2}\Vert \mathbf{r}-\mathbf{H}\mathbf{x}\Vert^2+{\text{Re}}\innerproduct{\mathbf{y}}{\mathbf{x}-\mathbf{z}}
  	\\&+\frac{\rho}{2}\Vert \mathbf{x}-\mathbf{z}\Vert^2
  	\end{split}\\  
  	\mathbf{z}^{t+1}:={}&\argmin_{\mathbf{z}} \alpha_1\sum_{j=1}^J w_j \Vert\mathbf{z}_j\Vert_2+{\text{Re}}\innerproduct{\mathbf{y}}{\mathbf{x}-\mathbf{z}}\nonumber\\&+\frac{\rho}{2}\Vert \mathbf{x}-\mathbf{z}\Vert^2\label{eq_18b}\\
  	\mathbf{y}^{t+1}:={}&\mathbf{y}^{t}+\rho(\mathbf{z}^{t+1}-\mathbf{x}^{t+1})\label{eq_18c} 	
  \end{align}
\end{subequations}
   The three ADMM solutions are derived by following the similar procedure of ADMM for sparse group LASSO  in subsection \ref{JADD_SG_LASSO}
   \begin{subequations}   \label{eq_23set}
    \begin{align}	\small	
   	\mathbf{x}^{t+1}:={}& (\mathbf{H}^{\rm H}\mathbf{H}+\rho \mathbf{I})^{-1}(\mathbf{H}^{H}\mathbf{y}+\rho(\mathbf{z}^k-\mathbf{u}^k))\label{eq_22a}\\
   	\mathbf{z}_j^{t+1}:={}&\frac{(\mathbf{x}_j^{t+1}+\mathbf{u}_{1j}^{t})}{\Vert \mathbf{x}_j^{t+1}+\mathbf{u}_{1j}^{t}\Vert}\bigg( \Vert \mathbf{x}_j^{t+1}+\mathbf{u}_{1j}^{t}\Vert-\frac{\alpha_1w_j^t}{\rho} \bigg)_+,\quad\forall j\label{eq_22b}\\
   	\mathbf{u}^{t+1}:={}&\mathbf{u}^t+(\mathbf{x}^{t+1}-\mathbf{z}^{t+1}) \label{eq_22c} 
   \end{align}
\end{subequations}
\subsection{JADD for frame-wise dynamic block sparsity model (\ref{eq_Dy})}
The sparse group LASSO and group LASSO problems are formulated for multi-user detection in one-shot grant-free SCMA and DCMA systems, respectively in subsection~\ref{sub_SGLASSO}. Further,  ADMM based detection algorithm is developed to recover the  activity and data jointly for one-shot block sparsity model. This section deals the JADD in the more practical scenario. Thus, a frame-wise dynamic system model introduced in Subsection~\ref{sub_c} considered for the JADD. The ADMM based detection algorithm is adapted for the dynamic system model in Subsection~\ref{sub_c}. However, (\ref{eq_LASS1}) and (\ref{eq_LASS_DCMA}) doesn't requires the prior information of active user set and signal estimate due to  one-shot transmission.
A convenient approach for utilizing prior knowledge of both the support and the signal in the sparse recovery problem is proposed in~\cite{prior_Lu}. This approach is later adapted to incorporate the temporal correlation between consecutive time slots in a frame-wise model as prior information, enhancing the performance of MUD~\cite{Cirik}.
  The prior information aided group LASSO is  formulated to solve the multi-user detection problem of \textbf{Case-2} exits in dynamic scenario, for DCMA systems. The similar procedure is applicable to sparse group LASSO  detection problem given in (\ref{eq_LASS1}). 

The active user sets between the time slots varies randomly in the dynamic system model. When we do the multi-user detection via Algorithm  for each time slot independently, leads to degradation in the performance. A natural correlation exists between the active user sets among  the adjacent  time slots in the frame. Hence, this can be incorporated into the BCS problem to enhance the performance of the proposed MUD.

There are two cases exists in the dynamic system model discussed in  Subsection~\ref{sub_c}. To recall, the first case is where the user activity changes at a very fast rate, and the second case is where the user activity changes at a slow or moderate rate. This leads to two different approaches for JADD in dynamic system model.\\
\textbf{Case-1:} This can be termed as the worst case scenario, occurs only when the user activity changes at a very fast rate. It implies, the activity of the users does not shared by adjacent time slots.
As a result, there is no  temporal correlation exists between the active user sets  among consecutive time slots in a frame. Therefore, the observation in each time slot can be processed independently as  follows
\begin{align}
	\min_{\mathbf{x}^{[l]}}\quad \Vert \mathbf{r}^{[l]}-\mathbf{H}^{[l]}\mathbf{x}^{[l]}\Vert^2+\underbrace{\alpha_1\sum_{j=1}^Jw_j\Vert\mathbf{x}_j^{[l]}\Vert_2}_{\text{user wise sparsity}}\nonumber\\ ,l=1,2,\hdots,L \label{eq_LASS_ind}
\end{align}
\textbf{Case-2:} Assuming that user activity changes slowly within a frame, there is a high probability that common active users will appear across adjacent time slots. This results in a temporal correlation between the  active user set in consecutive time slots. As a result, the ADMM aided multi-user detection for JADD can be performed in two key steps as discussed below.

\textbf{Step-i:} First apply Algorithm~\ref{alg1} independently to (\ref{eq_LASS_ind}) and compute active user set $\Lambda_{\rm p}^{[l]}$ and signal estimate $\hat{\mathbf{x}}_p^{[l]}$ on each time slot, $l=1,2,\hdots,L$. We can easily extract the codeword or signal  index value  $I_{\rm p}^{[l]}$ form $\hat{\mathbf{x}}_p^{[l]}$. Note that, $\hat{\mathbf{x}}_p^{[l]}$ contains value of the signal estimate and  $I_{\rm p}^{[l]}$ contains index values of the signals transmitted on $l$th time slot.  The subscript p indicates the prior information  for \textbf{Step-ii}.

After performing  \textbf{Step-i}, compute the common support set between the time slots as follows
\begin{align}
	\Lambda_{\rm p}^{[l]}\cap\Lambda_{\rm p}^{[l-1]}=&\emptyset, \quad  \hat{\mathbf{X}} \in  \text{Case-1}\nonumber\\
	\neq & \emptyset,\quad\hat{\mathbf{X}}\in  \text{Case-2},\quad \forall l \label{step-2}
\end{align}
Recall $\hat{\mathbf{X}}=[\hat{\mathbf{x}}^{[1]},\hat{\mathbf{x}}^{[2]},\hdots,\hat{\mathbf{x}}^{[L]}]$. If $\hat{\mathbf{X}}\in  \text{Case-1}$, which is under worst case scenario. Thus, there are no common users found between the adjacent time slots. Hence, the active user set $\Lambda_{\rm p}^{[l]}$ and signal estimate $\hat{\mathbf{x}}^{[l]}$ computed independently from (\ref{eq_LASS_ind}),  for $l=1,2,\hdots,L$ are the final estimates for the \textbf{Case-1 }. 

If $\hat{\mathbf{X}}\in  \text{Case-2}$, it leads to perform \textbf{Step-ii} in two ways as follows 
\begin{align}
	I_{\rm p}^{[l]}\cap I_{\rm p}^{[l-1]}=&\emptyset, \quad \hat{\mathbf{X}} \in  \text{Case-2a}\nonumber\\
	\neq & \emptyset,\quad \hat{\mathbf{X}} \in  \text{Case-2b},\quad \forall l. \label{step-2_case2b}
\end{align}

\textbf{Step-ii:}
This temporal correlation exists between the active user sets and the signal information between the time slots in a frame can be exploited in two ways to enhance performance. As discussed in Subsection~\ref{sub_c}, the JADD of two sub-cases within Case-2 as outlined below:\\
\textbf{Case-2a: (Symbol duration<time slot) }\\
In this, we assume symbol duration is less than the time slot. Thus, the symbols transmitted by the  active user in multiple adjacent time slots are different. The temporal correlation exits only in active user sets between the time slots and symbols transmitted by them within the frame are independent. Therefore, the active user set estimated in the previous time slot can be exploited to estimate the active user set in the present time slot.

The active user sets are temporally correlated and also dynamic throughout the frame. The user can come and go at any time slot within the frame. Hence, blindly utilizing the active set in the previous time slot leads to degradation in the performance. Instead, we can compute the partial information of $l$th time slot from the $(l-1)$th time slot.
 It is to be noted that, neither active user sets nor signal values are known at the receiver to exploit the temporal correlation. Hence, the proposed algorithm utilizes the computed support set and signal values in \textbf{Step-i} as partial information for \textbf{Step-ii}. \\

 \textbf{Step-i} computes the support set and signal estimates for $l=1,2,\hdots,L$.  \textbf{Step-ii} in Case-2a exploits the partial active user set computed from the  \textbf{Step-i} as a prior information.

If  $\hat{\mathbf{X}} \in  \text{Case-2a}$, there exists some common active users between the time slots. Then algorithm proceeds further to incorporate the  partial information computed in \textbf{Step-i} as prior information in \textbf{Step-ii} as follows
\begin{align}
	\min_{\mathbf{x}^{[l]}}\quad \Vert \mathbf{r}^{[l]}-\mathbf{H}^{[l]}\mathbf{x}^{[l]}\Vert^2+\alpha_1\sum_{j=1}^Jw_{j,p}^t w_{j,d}\Vert\mathbf{x}_j^{[l]}\Vert_2\nonumber\\ ,l=1,2,\hdots,L \label{eq_step2}
\end{align} 
where $w_{j,p}^l$ is the weight associated with the prior information of the $j$th user in $l$th time slot  and $w_{j,d}$ is the weight associated with democratic penalization.
$	w_{j,p}^t$ can be defined as
\begin{align}
	w_{j,p}^l=&0, \quad j\in q^{[l]}\nonumber\\
	=& 1, \quad   j\not\in q^{[l]}
\end{align}
where  $ q^{[l]}=\Lambda^{[l-1]}\cap \Lambda_{\rm p}^{[l]}$ is the temporally correlated active user set between $(l-1)$ and $l$th time slots. $q^{[l]}$ improves the reconstruction quality on the $l$th time slot.  Hence, $ q^{[l]}$ indicates the quality support set for the $l$th time slot evaluated from the  prior information. The prior information consists of the active user set taken  from the previous time slot  $\Lambda^{[l-1]}$ and the  active user set taken from \textbf{Step-i}  of the current time slot  $ \Lambda_{\rm p}^{[l]}$. 
$\Lambda^{[l-1]}$ is the active user set of the previous time slot, computed by applying ADMM algorithm on (\ref{eq_step2}), and $\Lambda_{\rm p}^{[l]}$ indicates the prior active user set of present time slot computed by applying Algorithm~\ref{alg1} on (\ref{eq_LASS_ind}) independently for $l=1,2,\hdots,L$. $	w_{j,p}^t$ ensures the block-wise sparse vector $\mathbf{x}$, which is sparsest  outside the set $q^{[l]}$ in $l$th time slot. $w_{j,d}$ ensures the democratic penalization for the users, which are outside the quality support set $q^{[l]}$. Thus, the second term in (\ref{eq_step2}) doesn't penalize the active users whose locations are known i.e. quality set $q^{[l]}$.  
 
\textbf{Case-2b: (Symbol duration>time slot)}\\
 Let us  consider a case, when the symbol duration is larger than the time slot. Thus, with high probability the slowly varying active user transmits the same information across multiple adjacent time slots within the frame. Hence, the temporal correlation of both active user sets, as well as the signal estimates between the time slots further improves the reconstruction accuracy. The  JADD for \textbf{Case-2b}  follows the two step procedure of \textbf{Case-2a}. 
 
If $\mathbf{X}$ falls under \text{Case-2b}, the quality information of the present time slot includes both the partial support set and the signal estimate values for that set coming from the previous time slot. As a result, the group LASSO problem (\ref{eq_step2}) in \textbf{step-ii} of \textbf{Case-2a} is modified accordingly.
  By incorporating the partial signal information corresponding to the quality support set $q^{[l]}$, the  group LASSO problem for \text{Case-2b}  is as follows
 
 \begin{align}
 	\min_{\mathbf{x}^{[l]}}\quad \Vert \mathbf{r}^{[l]}-\mathbf{H}^{[l]}\mathbf{x}^{[l]}\Vert^2+\alpha_1\sum_{j=1}^Jw_{j,p}^l w_{j,d}\Vert\mathbf{x}_j^{[l]}\Vert_2 \nonumber \\+\frac{\mu^{l}}{2}\Vert \mathbf{x}^{[l]}-\hat{\beta}_{q^{[l]}}^{[l-1]} \Vert^2    ,l=1,2,\hdots,L \label{eq_Case2b}
 \end{align}
where $\hat{\beta}_{q^{[l]}}^{[l-1]} $ is the partial signal information of $l$th time slot taken from the $(l-1)$th time slot. It contains the non-zero values for the users of the quality support set, $q^{[l]}$. The non-zero values indicate the signal estimates taken from the previous time slot.  The third term in (\ref{eq_Case2b}) ensures the signal to be estimated is close enough to $\hat{\beta}_{q^{[l]}}^{[l-1]}$.

By following the similar procedure given in subsection~\ref{JADD_SG_LASSO}, 3-ADMM solutions derived  for the problem  (\ref{eq_Case2b})  as follows
\begin{subequations}   \label{case_2b_steps}
	\begin{align}	\small	
		\mathbf{x}^{[l],t+1}:={}& \big(\mathbf{H}^{[l]^{\rm H}}\mathbf{H}^{[l]}+(\mu^l+\rho) \mathbf{I}\big)^{-1}\big(\mathbf{H}^{[l]^{\rm H}}\mathbf{r}^{[l]}+\mu^l\hat{\beta}_{q^{[l]}}^{[l-1]}\nonumber\\&\hspace{3cm}+\rho(\mathbf{z}^{[l],t}-\mathbf{u}^{[l],t})\big)\label{eq_29a}\\
		\mathbf{z}_j^{[l],t+1}:={}&\frac{\mathbf{p}_j}{\Vert \mathbf{p}_j\Vert_2}\bigg( \Vert \mathbf{p}_j\Vert_2-\frac{\alpha_1w_{j,d}^t w_{j,p}^l}{\rho} \bigg)_+,\quad\forall j\label{eq_29b}\\		
		\mathbf{u}^{[l],t+1}:={}&\mathbf{u}^{[l],t}+(\mathbf{x}^{[l],t+1}-\mathbf{z}^{[l],t+1}), \label{eq_29c} 
	\end{align}
\end{subequations}
where, $\mathbf{p}_j=\mathbf{x}_j^{[l],t+1}+\mathbf{u}_{j}^{[l],t}$.

\section{Computational Complexity}
This section discuss the computational complexity of the proposed ADMM based detection algorithms given in Algorithm~\ref{alg1} and Algorithm~\ref{alg2}. Algorithm~\ref{alg2} is a extended version of Algorithm~\ref{alg1} and requires more number of computations than Algorithm~\ref{alg1} to address the practical issues in dynamic scenarios. The computational complexity depends on the number of   floating point operations (FLOPs) or complex multiplications.  The number of FLOPS required to perform these algorithms discussed as follows 

Algorithm~\ref{alg1} consists of three parts: pre-processing, iterative processing, and post-processing.  The pre-processing include the matrix multiplications and  matrix inverse operation such as $\mathbf{H}^H\mathbf{H}$, $\mathbf{H}^H\mathbf{r}$, and $(\mathbf{H}^H\mathbf{H}+2\rho I)^{-1}$. The size of the $\mathbf{H}$ matrix is $N_{\rm r}K\times JK$. The number of FLOPs required to perform these three operations are $(N_{\rm r}K)(KJ)^2$, $(N_{\rm r}K)(KJ)$, and $(KJ)^3$, respectively. The iterative processing steps are iteration dependent and need to be repeated in each iteration. The computations  in \textbf{Step:1} in (\ref{eq_22a}) include scalar multiplication with a vector of size $KJ\times 1$ and  the multiplication of a matrix of size $KJ\times KJ$ with a vector of size $KJ\times 1$. The number of FLOPs required to perform these computations are $(KJ)+(KJ)^2$. The \textbf{Step:2} in (\ref{eq_22b}) need to be performed for $j=1,2,\hdots,J$, and which  includes the multiplication of $K\times 1$ vector with a scalar. The number of FLOPs required to perform \textbf{Step:2} is $KJ$. The \textbf{Step-3} and the post-iteration processing of Algorithm~\ref{alg1} doesn't include any complex multiplications and have negligible complexity. Therefore, the total number of FLOPs required in the iterative processing of the Algorithm~\ref{alg1} are $(N_{\rm r}K)(KJ)^2+(KJ)^3+(N_{\rm r}K)(KJ)+T(2KJ+(KJ)^2)$.

The Algorithm~\ref{alg2} proposed for a frame-wise dynamic block-sparsity model needs to perform the Algorithm~\ref{alg1} for $l=1,2,\hdots,L.$ Algorithm~\ref{alg2} address the key issues in JADD of dynamic scenarios, and  performed in two steps. In which, \textbf{Step-i} performs independent JADD  and  \textbf{Step-ii} performs the JADD by incorporating the prior information computed from \textbf{Step-i}.
The number of FLOPs required to perform in \textbf{Step-i} of the Algorithm~\ref{alg2} are $L((N_{\rm r}K)(KJ)^2+(KJ)^3+(N_{\rm r}K)(KJ)+T((KJ)+(KJ)^2))$. The prior information aided ADMM iterations in \textbf{Step-ii} further improves the performance at the cost of increasing the complexity. 
 In \textbf{Step-ii }, the Algorithm~\ref{alg1} need to be performed second time for $l=1,2,\hdots,L$ by incorporating the prior information computed from   \textbf{Step-i }. 
It can be observed in Algorithm~\ref{alg2}, if $\hat{\mathbf{X}}\in \text{Case-1}$, then algorithm stops after \textbf{Step:i}. Hence, the complexity is same as the complexity of \textbf{Step:i}. If  $\hat{\mathbf{X}}\in \text{Case-2}$,  then algorithm performs \textbf{Step:i} and \textbf{Step:ii}. Hence, the total number of FLOPs for Algorithm~\ref{alg2}  are $2L((N_{\rm r}K)(KJ)^2+(KJ)^3+(N_{\rm r}K)(KJ)+T((KJ)+(KJ)^2))$. 

The Prior Information-Aided SP algorithm for JADD~\cite{prior} effectively addresses challenges in dynamic environments. In contrast, the Block Subspace Pursuit (BSP) algorithm tackles the JADD problem within a frame-wise static block-sparsity model~\cite{du_block}. The Prior Information-Aided BSP algorithm, applicable to the block-sparsity models discussed in Section~\ref{sec_II}, serves as a suitable baseline for comparison with the proposed method. The BSP algorithm follows several steps in its implementation, and its computational complexity is thoroughly analyzed in~\cite{du_block}. The number of floating point operations (FLOPs) required for the Oracle BSP algorithm is outlined as follows. \\
\textbf{support estimate:} The support estimate requires $(J(K^2N_{\rm r})+JK+K)$.\\
\textbf{least squares estimation:} The FLOPS required to perform LSE are $(\vert\tilde{T}\vert K)^3+2KN_{\rm r}(\vert\tilde{T}\vert K)^2$, where $\tilde{T}$ is the merged support set in support estimation.\\
\textbf{ Support pruning:} The support pruning performs $(KJ+J)$ FLOPs. \\
\textbf{Signal estimation:} The signal estimation performs $(J_{\rm a} K)^3+2KN_{\rm r}(J_{\rm a} K)^2$ FLOPs, where $J_{\rm a}$ is the size of true support set.\\
\textbf{Residue computation:} Residue computation requires $(KN_{\rm r})(KJ)$ FLOPs.\\
The above steps need to be repeated in each iteration of the BSP until it converges. The maximum number of iterations  of BSP algorithm is indicated as  $T_{\rm BSP}$.
The computational complexity of single iteration of the iterative BSP algorithm $(J(K^2N_{\rm r})+JK+K)+((\vert\tilde{T}\vert K)^3+2KN_{\rm r}(\vert\tilde{T}\vert K)^2)+(KJ+J)+((J_{\rm a} K)^3+2KN_{\rm r}(J_{\rm a} K)^2)+(KN_{\rm r})(KJ).$
\section{Simulations and discussions}
The numerical results of the proposed detection algorithm for various CB based grant-free system models are discussed in this section.
\subsection{SER performance: One-shot block sparsity model}\label{sim_A}
\begin{figure*}[htb!]
	\centering	
	\begin{subfigure}[h]{0.4\linewidth}			
		\centering
		\includegraphics[width=\textwidth]{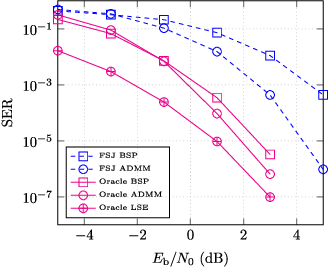} 
		\caption{\footnotesize{DCMA system with  $J$=8 and $K$=4  }}     
		\label{Fig. 1(a)}
	\end{subfigure}\hspace{2em}  
	\begin{subfigure}[h]{0.4\linewidth}\label{dcma_model}
		
		\centering
		\includegraphics[width=\textwidth]{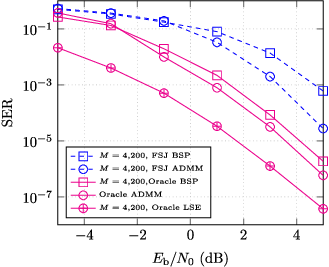}      
		\caption{\footnotesize{SCMA system with  $J$=8 and $K$=4 }}
		\label{Fig. 1(b)}
	\end{subfigure}
	\caption{SER performance of one shot static CD-NOMA systems.}
	
	\label{fig1}
\end{figure*}

This section presents the simulation results of the proposed detection algorithm for the one-shot block sparsity model in the uplink scenario. Figure \ref{fig1} compares the SER performance of different detection schemes for both DCMA and SCMA systems. Figure \ref{Fig. 1(a)} shows the SER performance curves for DCMA systems. The Oracle least squares estimation (Oracle LSE) is used as a benchmark for comparison. Oracle LSE assumes the true support set is known to the receiver and performs LSE on this known support set to estimate the signal information. The Oracle ADMM assumes the sparsity level $S_{\rm l}$ is known to the receiver and performs ADMM iterations, as outlined in Algorithm~\ref{alg1}, to jointly estimate activity and signal information. From the figure, it is evident that as the signal-to-noise ratio (SNR) increases, the performance of Oracle ADMM approaches that of Oracle LSE. The FSJ-aided ADMM algorithm, on the other hand, performs joint activity detection and decoding (JADD) without knowing $S_{\rm l}$ in advance, leading to some performance degradation. However, the adapted FSJ technique still provides acceptable performance. Meanwhile, the FSJ-aided block sparsity subspace pursuit algorithm (BSA) performs worse than the FSJ-aided ADMM.The ADMM-based detector achieves an SNR gain of approximately 1.2 dB compared to the BSP algorithm
Figure \ref{Fig. 1(b)} shows the SER performance comparison for SCMA systems, where similar observations as those for DCMA systems apply.
\subsection{SER performance: Frame-wise dynamic block sparsity model}\label{sim_B}
\begin{figure*}[htb!]
	\centering
	
	\begin{subfigure}[h]{0.4\linewidth}\label{scma_model}
		
		\centering
		\includegraphics[width=\textwidth]{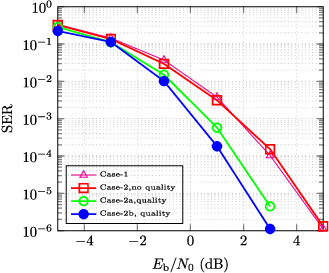}  
		\caption{\footnotesize{Various cases in dynamic system  }}    
		\label{Fig. 2(a)}
	\end{subfigure}\hspace{2em}  
	\begin{subfigure}[h]{0.4\linewidth}\label{dcma_model}
		
		\centering
		\includegraphics[width=\textwidth]{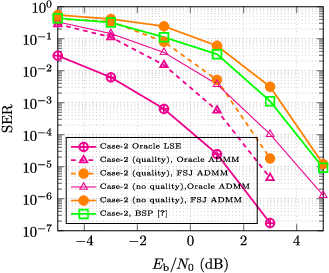}     
		\caption{\footnotesize{Oracle ADMM vs. adaptive ADMM }} 
		\label{Fig. 2(b)}
		
	\end{subfigure}
	
	\caption{SER performance: Frame-wise dynamic block-sparsity model.}
	\label{fig_2}
\end{figure*}

This subsection presents various numerical experiments conducted on the proposed frame-wise dynamic block sparsity model. The advantages of several practical use cases and their detection methodologies, as proposed in previous sections, are highlighted here. Fig.~\ref{fig_2} shows the SER performance comparison for different scenarios in an uplink DCMA system under dynamic conditions. In Fig.~ \ref{Fig. 2(a)}, the SER performance curves for Algorithm~\ref{alg2} across various cases are illustrated. It is evident that the performance of Case-1 is inferior to the other curves. This is because Case-1 lacks temporal correlation between the active user sets across time slots, resulting in no available quality information for collaborative time-slot processing. In this paper, Case-1 is referred to as the worst-case scenario. On the other hand, Case-2, while having quality information, performs similarly to Case-1 if the quality information is not exploited, making its performance comparable to the worst-case scenario. Additionally, Fig. \ref{Fig. 2(a)} illustrates the benefits of utilizing quality information in terms of performance. In Case-2a, the quality information is represented by common active user sets across time slots. With this quality information, Case-2a achieves an SNR gain of approximately 1.2 dB at an SER of $10^{-4}$ compared to Case-2 without quality information. Case-2 without quality information is a blind exploitation of temporal correlation, which is equivalent to the idea in ~\cite{Cirik}. Case-2b, on the other hand, incorporates quality information in the form of both common active user sets and their signal details. As a result, the increased quality information in Case-2b leads to superior performance compared to all other cases.

Figure \ref{Fig. 2(b)} compares the SER performance of different detection schemes for the Case-2 scenario. This includes the performance of Oracle ADMM and FSJ-aided ADMM, both with and without quality information. It is evident that in both ADMM schemes, the inclusion of quality information significantly enhances performance compared to the versions without it. The SER performance of the two ADMM schemes with quality information is much closer to the benchmark performance compared to their counterparts without quality information. Furthermore, FSJ-aided ADMM with quality information achieves an SNR gain of nearly 2 dB at an SER of $10^{-5}$ over FSJ-aided ADMM without quality information. Further, Figure \ref{Fig. 2(b)} includes the performance of existed prior-information aided adaptive SP (PIA-ASP) in \cite{prior} algorithm. The performance of PIA-ASP is almost similar to the no quality aided FSJ ADMM.The quality-aided FSJ ADMM outperforms the PIA-ASP scheme by nearly 1.9 dB in SNR gain at an SER of approximately $10^{-5}$. While the PIA-ASP scheme enhances detector performance by incorporating the quality parameter ($\rm s_{\rm p}$), it assumes that this parameter is inaccurately known to the receiver, which limits its effectiveness. Additionally, the $\rm s_{\rm p}$ in PIA-ASP represents the size of the common active users between two consecutive time-slots. In contrast, the proposed MUD leverages quality information in the form of both partial active user support sets $q^{[l]}$ and signal information $\hat{\beta}_{ q^{[\rm l]}}$. Rather than assuming the quality information is known to the receiver, it is computed from \textbf{Step-i}. Figure \ref{fig2} underscores the significance of using quality information in ADMM iterations to enhance reliability in practical scenarios.
\subsection{SER performance : channel estimations errors (CEEs)}
\begin{figure}[htb!]		
	\centering
	\includegraphics[scale=1.2]{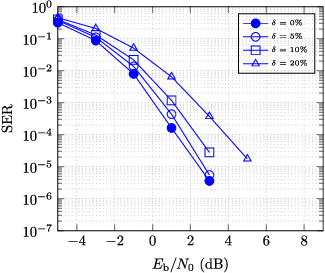}
\caption{SER performance: imperfect CSI scenario}
\label{fig.CEE}
\end{figure}

The proposed algorithm considers perfect channel state information (CSI) available at the receiver. The extensive simulations in this section show the effectiveness of the ADMM based MUD for JADD in various scenarios. This section highlights the performance of the proposed detector against channel estimations errors (CEEs). In practice, the channel information is not available to the receiver and needs to be estimated prior to the data detection. Thus, the perfect channel estimation is not possible due to the channel estimation errors occurs in the estimation process. As a result, imperfect CSI is often available to the receiver. Thus, it is important to analyze the performance of the detector in the presence of CEEs. A standard model is used to introduce the CEEs as follows
\begin{align}\label{eq_CEE}
\widetilde{\mathbf{H}}=\mathbf{H}+\delta\mathbf{\Omega}
\end{align}
 where $\mathbf{\Omega}$ indicates the CEE assumed to be uncorrelated with $\mathbf{H}$. The entries of $\mathbf{\Omega}$ are i.i.d complex Gaussian random variables with zero mean and unit variance, denoted as $\mathcal{CN}(0,1)$. The parameter $\delta$ determines  the variance of $\mathbf{\Omega}$. Fig. \ref{fig.CEE} show the performance of the proposed detector for the imperfect channel matrix given in (\ref{eq_CEE}) with $\delta=0\%$, $\delta=5\%$, $\delta=10\%$, and $\delta=20\%$. Observe  that the ADMM based MUD is robust against CEEs. This results show that the proposed detector gives acceptable performance in imperfect CSI scenarios.

\subsection{ Convergence}
\begin{figure*}[h]
	\centering
	\caption{Residual Convergence.}
	\begin{subfigure}[h]{0.4\linewidth}\label{scma_model}
		\caption{\footnotesize{Primal residual convergence  }}
		\centering
		\includegraphics[width=\textwidth]{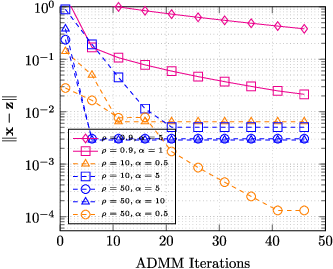}      
		\label{Fig.prim_res(a)}
	\end{subfigure}\hspace{2em}  
	\begin{subfigure}[h]{0.4\linewidth}\label{dcma_model}
	
		\centering
		\includegraphics[width=\textwidth]{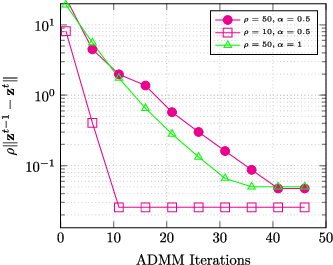}      
		
			\caption{\footnotesize{Dual residual convergence}}
				\label{Fig. dual_res(b)}
	\end{subfigure}
\label{fig_res}
	\vspace{-2\baselineskip}	
\end{figure*}

This subsection presents an empirical analysis of the residual convergence of the proposed ADMM-based detection algorithm, demonstrating the algorithm's convergence behavior. A more rigorous theoretical analysis is provided in the Appendix, establishing the guaranteed theoretical convergence properties of the ADMM-based detection algorithm. Additionally, the empirical observations on convergence aid in selecting key ADMM parameters, specifically $\rho$ and $\alpha_1$ in (\ref{eq_LASS_DCMA}), and $T$ in Algorithm~\ref{alg1}. These parameters significantly influence both the convergence speed and the recovery performance of the algorithm. In this work, the ADMM parameters are chosen based on the empirical observations shown in Fig.~\ref{fig_res}. However, optimal parameter selection could further enhance the algorithm's speed and accuracy. Achieving this optimal selection is a complex task and falls beyond the scope of this work.

he necessary and sufficient optimality conditions for the ADMM detection problem, based on primal and dual feasibility, can be applied~\cite{boyd}. This leads to two key quantities for checking residual convergence: the primal residual and the dual residual. For the problem (\ref{eq_LASS_DCMA}), the primal residual is given by $\mathbf{r}_p^{(t)} = \mathbf{x}^{t} - \mathbf{z}^{t}$ and the dual residual by $\mathbf{r}_d^{(t)} = \rho(\mathbf{z}^t - \mathbf{z}^{t+1})$. These residuals are also used as stopping criteria in Algorithm~\ref{alg1}. When their values drop below the feasibility tolerances, the ADMM terminates, indicating convergence.

Figs. \ref{Fig.prim_res(a)} and \ref{Fig. dual_res(b)} illustrate the primal and dual residual convergence for Algorithm~\ref{alg1}, showing that the proposed ADMM-based detection algorithm converges in a few tens of iterations. From Fig.~\ref{fig_res}, it's clear that the parameters $\rho$ and $\alpha_1$ significantly affect the convergence speed. A larger $\rho$ imposes a heavier penalty on primal feasibility violations, leading to smaller primal residuals. Conversely, smaller $\rho$ values result in smaller dual residuals. However, excessively large or small values of $\rho$ are detrimental to both primal and dual residual convergence. On the other hand, $\alpha_1$ in (\ref{eq_LASS_DCMA}) controls the sparsity of the solution in (\ref{eq_22b}). Higher $\alpha_1$ values intensify the $\rm L_1$ penalty, producing a highly sparse or even all-zero solution for $\mathbf{z}^{(t)}$, making the dual residual $\mathbf{r}_d^{(t)}$ zero. This is why Fig.\ref{Fig. dual_res(b)} excludes curves for $\alpha_1 > 1$. The selection of $\rho$ and $\alpha_1$ is interdependent, and achieving a balance between the two is crucial for convergence speed and reconstruction accuracy. In summary, moderate values of $\rho$ and smaller $\alpha_1$ values are preferable for the proposed ADMM detection algorithm. Fig.\ref{fig_res} shows that with $\rho = 10$ and $\alpha_1 = 0.5$, the algorithm converges quickly in both primal and dual residuals. The proposed algorithm is converges to modest accuracy almost in ten iterations. This is sufficient to recover the signal with acceptable SER performance discussed in subsections~\ref{sim_A}\&~\ref{sim_B}.

\subsubsection{Convergence of \textbf{Algorithm 1}}
The convergence of ADMM algorithm for convex problems is quiet extensively investigated. It is worth mentioning that the ADMM is known to be convergent under mild selection of parameter $\rho$. 
{\color{black}
	\begin{theorem}\label{th1}
		Suppose $\rho$ is sufficiently large and the conditions $\rho>0$, $\rho(\rho+\lambda_{\rm min}(\mathbf{H}^H\mathbf{H}))\geq 2 \lambda_{\rm max}^2(\mathbf{H}^H\mathbf{H})$  and $\rho>\lambda_{\rm max}(\mathbf{H}^H\mathbf{H})$ are satisfied. Then the sequence $\{\mathbf{x}^t,\{\mathbf{z}_{j}^t\}_{j=1}^J,\mathbf{y}^t\}$ generated by ADMM steps in (\ref{eq_23set})  is convergent as given below
		{\small\begin{align}
				\lim_{t\to\infty} \mathbf{x}^t=\mathbf{x}^\star,\quad
				\lim_{t\to\infty} \mathbf{z}_j^t=\mathbf{z}_j^\star,\forall j\quad
				\lim_{t\to\infty} \mathbf{y}^t=\mathbf{y}^\star.
		\end{align}}
		\begin{proof}
			Provided in Appendix.
		\end{proof}    
		\textbf{Remark 1:} The above theorem proves the convergence of the proposed ADMM algorithm for  Group LASSO problem formulated for DCMA systems. The derivation and the conditions on $\rho$ given for problem ~(\ref{eq_LASS_DCMA})holds for problem ~(\ref{eq_LASS1}).		
		The theorem proves that the proposed ADMM detection algorithm for Group LASSO based detection problem is guaranteed to converge to some stationary point under mild conditions on the selection of the ADMM penalty parameter $\rho$. The parameters are selected based on the given channel matrix $\mathbf{H}$.

		
\end{theorem}}

{\color{black}\appendix
	\subsection{Convergence of ADMM algorithm on  problem~(\ref{eq_LASS_DCMA})}\label{sub_app}
	The following assumptions are made to analyze the proposed ADMM-based detection algorithm, and are subsequently validated through Lemmas..\\
	\textbf{Assumptions:}\\
	\begin{itemize}
		\item The function $f$ is Lipschitz continuous, then there exists a positive constant $m$ such that
		$$\Vert \nabla_{\mathbf{x}} f(\mathbf{x}^{t+1})-\nabla_{\mathbf{x}} f(\mathbf{x}^{t})\Vert\leq
		 m\Vert \mathbf{x}^{t+1}-\mathbf{x}^{t}\Vert$$
		 where $f(\mathbf{x})=\frac{1}{2}\Vert \textbf{r}-\mathbf{H}\mathbf{x}\Vert_2^2$.
		 \item The parameter $\rho$ is chosen  positive and large enough such that the sub-problems related to $x$ and $z$ are strongly convex with modulus $\gamma_1$ and $\gamma_2$, respectively, where $\rho\gamma_1>2m^2$ and $\rho\geq m$. It implies  that $\mathcal{L}\left(\{\mathbf{x}^{t},\{\mathbf{z}_j^{t}\},\mathbf{y}^t\right)$ always decreases and lower bounded as follows
		 \begin{align*}
		 	\mathcal{L}\left(\mathbf{x}^{t},\{\mathbf{z}_j^{t}\},\mathbf{y}^t\right)&>\mathcal{L}\left(\{\mathbf{x}^{t+1},\{\mathbf{z}_j^{t+1}\},\mathbf{y}^{t+1}\right)\\
		 \lim_{t\to\infty}\mathcal{L}\left(\mathbf{x}^{t},\{\mathbf{z}_j^{t}\},\mathbf{y}^{t}\right) &> -\infty
		 \end{align*}
	 where $	\mathcal{L}\left(\{\mathbf{x}^{t},\{\mathbf{z}_j^{t}\},\mathbf{y}^t\right)$ is the augmented Lagrangian function of problem (\ref{eq_LASS_DCMA}) is given by
	 \begin{align}\label{eq_aug_DCMA}
	 	\begin{split}
	 		\mathcal{L}\left(\mathbf{x},\{\mathbf{z}_j\}_{j=1}^J,\mathbf{y}_1\right)=	\frac{1}{2}\Vert \mathbf{r}-\mathbf{H}\mathbf{x}\Vert_2^2+\alpha_1\sum_{j=1}^Jw_j\Vert\mathbf{z}_j\Vert_2\\ + {\text{Re}}\innerproduct{\mathbf{y}_1}{\mathbf{x}-\mathbf{z}}
	 		+\frac{\rho}{2}\Vert \mathbf{x}-\mathbf{z}\Vert_2^2
	 	\end{split}
	 \end{align}

	\end{itemize}

The following three lemmas are used to prove \textbf{Theorem 1} according to \cite{Hong_SIAM}. \textbf{ Lemma 1} proves the function,$f(\mathbf{x})$ is Lipschitz continuous  w.r.t. constant $m=\lambda_{\rm max}(\mathbf{H}^H\mathbf{H})$\cite{Zhang_Quan}, it shows the difference between the dual variables in two consecutive iterations bounded above by that of the primal variables. \textbf{Lemma 2} derives conditions on penalty parameters under which augmented Lagrangian function $\mathcal{L}$ (\ref{eq_aug_DCMA}) always decreases. \textbf{Lemma 3} proves that there exists a lower bound on $\mathcal{L}$ with proper selection of penalty parameters. The proofs of Lemmas are provided below.
\begin{lemma} The following inequality holds for the proposed Algorithm~\ref{alg1}
\begin{align*}\tiny
    \Vert \nabla_{\mathbf{x}} f(\mathbf{x}^{t+1})-\nabla_{\mathbf{x}} f(\mathbf{x}^{t})\Vert^2\leq\lambda_{\rm max}^2(\mathbf{H}^H\mathbf{H})\Vert \mathbf{x}^{t+1}- \mathbf{x}^{t}\Vert^2.
\end{align*}
\vspace{-0.3in}
\begin{proof}
 Since $\mathbf{x}^{t+1}$ is a minimizer of (\ref{eq_18a}), the optimality condition is as follows \vspace{-1mm}
 \begin{align}\small
 \nabla_{\mathbf{x}} f(\mathbf{x}^{t+1})&+(\mathbf{y}^{t}+\rho(\mathbf{x}^{t+1}-\mathbf{z}^{t+1}))= 0\nonumber\\
   \text{From}\hspace{2mm}(\ref{eq_22c}), \quad
 & \nabla_{\mathbf{x}}  f(\mathbf{x}^{t+1})=-\mathbf{y}_{1}^{t+1} \label{ap.01}   
 \end{align}
 From the Lagrangian mean value theorem, \vspace{-2mm}
 \begin{align*} \small
     \frac{\nabla_{\mathbf{x}} f(\mathbf{x}^{t+1})-\nabla_{\mathbf{x}} f(\mathbf{x}^{t})}{\mathbf{x}^{t+1}-\mathbf{x}^{t}}= \nabla_{\mathbf{x}}^2 f(\mathbf{x})\nonumber\\
     \nabla_{\mathbf{x}}^2 f(\mathbf{x})
       =\mathbf{H}^H\mathbf{H}
      \preceq \lambda_{\rm max}(\mathbf{H}^H\mathbf{H}) I      
 \end{align*} 
       \begin{align} \small              
      \begin{split}
      \Vert \nabla_{\mathbf{x}} f(\mathbf{x}^{t+1})-\nabla_{\mathbf{x}} f(\mathbf{x}^{t})\Vert^2
      \leq \lambda_{\rm max}^2(\mathbf{H}^H\mathbf{H})\Vert \mathbf{x}^{t+1}- \mathbf{x}^{t}\Vert^2 
      \end{split}\label{ap.02}
        \end{align}
         From (\ref{ap.01}), \vspace{-2mm}
        \begin{align}\small
      \begin{split}
       \Vert \mathbf{y}^{t+1}-\mathbf{y}^{t}\Vert^2\leq\lambda_{\rm max}^2(\mathbf{H}^H\mathbf{H})\Vert \mathbf{x}^{t+1}- \mathbf{x}^{t}\Vert^2,\forall j.
        \end{split}\label{ap.05}               
  \end{align}   \end{proof}
\end{lemma}
\vspace{-2mm}
\begin{lemma}
    For the augmented Lagrangian function (\ref{eq_aug_DCMA}), the following holds true
    \begin{align} \small
        &\mathcal{L}\left(\mathbf{x}^{t},\{\mathbf{z}_j^{t}\},\mathbf{y}^t\right)- \mathcal{L}\left(\mathbf{x}^{t+1},\{\mathbf{z}_j^{t+1}\},\mathbf{y}^{t+1}\right) \nonumber \\
       & \geq\left( \frac{\gamma_1}{2} -\frac{\lambda_{\rm max}^2(\mathbf{H}^H\mathbf{H})}{\rho}  \right)\Vert \mathbf{x}^{t+1}-\mathbf{x}^{t}  \Vert^2\nonumber \\
       &+\sum_{j=1}^{J} \frac{\gamma_2}{2}\Vert \mathbf{z}^{t+1}-\mathbf{z}^{t}\Vert^2.
         \label{ap.1}
      \end{align}   
    \begin{proof}    
    \begin{align}  \small
    \begin{split}\label{eq_34}
         \mathcal{L}\left(\mathbf{x}^{t+1},\{\mathbf{z}_j^{t+1}\},\mathbf{y}^{t+1}\right) - \mathcal{L}\left(\mathbf{x}^{t},\{\mathbf{z}_j^{t}\},\mathbf{y}^{t}\right) \\=
     \underbrace{\left( \mathcal{L}\left(\mathbf{x}^{t+1},\{\mathbf{z}_j^{t+1}\},\mathbf{y}^{t+1}\right) - \mathcal{L}\left(\mathbf{x}^{t+1},\{\mathbf{z}_j^{t+1}\},\mathbf{y}^{t}\right)\right)}_{\text{term-1}}\\
         +\underbrace{\mathcal{L}\left(\mathbf{x}^{t+1},\{\mathbf{z}_j^{t+1}\},\mathbf{y}^{t}\right)- \mathcal{L}\left(\mathbf{x}^{t},\{\mathbf{z}_j^{t}\},\mathbf{y}^{t}\right)}_{\text{term-2}}
    \end{split}
 \end{align}


             \vspace{-0.1in}
             From (\ref{eq_22c}) and (\ref{ap.05}), the  term-1 in (\ref{eq_34}) is simplified as             
             \begin{align}\small
             \begin{split}  
           \mathcal{L}\left(\mathbf{x}^{t+1},\{\mathbf{z}_j^{t+1}\},\mathbf{y}^{t+1}\right) - \mathcal{L}\left(\mathbf{x}^{t+1},\{\mathbf{z}_j^{t+1}\},\mathbf{y}^{t}\right)\\ =\frac{1}{\rho}\Vert \mathbf{y}^{t+1}-\mathbf{y}^{t} \Vert^2 \leq \frac{1}{\rho} \lambda_{\rm max}^2(\mathbf{H}^H\mathbf{H})\Vert \mathbf{x}^{t+1}-\mathbf{x}^{t}\Vert^2
           \end{split}\label{ap.2}
            \end{align}

The term-2 in~(\ref{eq_34}) is given by
  \begin{align}\small
  \begin{split}
      \mathcal{L}\left(\mathbf{x}^{t+1},\{\mathbf{z}_j^{t+1}\},\mathbf{y}^{t}\right)- \mathcal{L}\left(\mathbf{x}^{t},\{\mathbf{z}_j^{t}\},\mathbf{y}^{t}\right) \\=\mathcal{L}\left(\mathbf{x}^{t+1},\{\mathbf{z}_j^{t+1}\},\mathbf{y}^{t}\right)- \mathcal{L}\left(\mathbf{x}^{t+1},\{\mathbf{z}_j^{t}\},\mathbf{y}^{t}\right) \\
      +\mathcal{L}\left(\mathbf{x}^{t+1},\{\mathbf{z}_j^{t}\},\mathbf{y}^{t}\right)- \mathcal{L}\left(\mathbf{x}^{t},\{\mathbf{z}_j^{t}\},\mathbf{y}^{t}\right)  \\  
     \overset{(a)}{\leq}   \sum_{j=1}^{J}\innerproduct{\varsigma_{\mathbf{z}_j^{t+1}}}{\mathbf{z}_j^{t+1}-\mathbf{z}_j^{t}}-\frac{\gamma_2}{2}\Vert\mathbf{z}_j^{t+1}-z_j^t\Vert_2^2\\+\innerproduct{\nabla_x \mathcal{L}\left(\mathbf{x}^{t+1},\{\mathbf{z}_j^{t}\},\mathbf{y}^{t}\right)}{\mathbf{x}^{t+1}-\mathbf{x}^t}-\frac{\gamma_1}{2}\Vert \mathbf{x}^{t+1}-\mathbf{x}^t\Vert_2^2  
      \\
      \overset{(b)}{\leq} -\frac{\gamma_1}{2}\Vert  \mathbf{x}^{t+1}-\mathbf{x}^{t} \Vert^2-\sum_{j=1}^{J}\frac{\gamma_2}{2}\Vert  \mathbf{z}^{t+1}-\mathbf{z}^{t} \Vert^2
      \end{split}\label{ap.3}
  \end{align} 
where (a) uses the fact that $\mathcal{L}\left(\mathbf{x},\{\mathbf{z_j}\},\mathbf{y}\right)$ is strongly convex \cite{boyd2004convex} w.r.t. each $\mathbf{x}$ and $\mathbf{z}$, with modulus $\gamma_1=\rho+\lambda_{min}(\mathbf{H}^H\mathbf{H})$ and $\gamma_2=\rho$, respectively.
 It is neccesaty that $\rho>0$ to hold the strong convexity of $\mathcal{L}\left(\mathbf{x},\{\mathbf{z}_j\},\mathbf{y}\right)$ w.r.t variable $\mathbf{z}_j$. 
  In (b) we have used the optimality conditions of sub-problems (\ref{eq_18a}) and (\ref{eq_18b}) given by $\varsigma_{\mathbf{z}_j^{t+1}}\in \partial_{\mathbf{z}_j} \mathcal{L}\left(\mathbf{x}^{t+1},\{\mathbf{z}_j^{t+1}\},\mathbf{y}^{t}\right)\in 0$ and $\nabla_x \mathcal{L}\left(\mathbf{x}^{t+1},\{\mathbf{z}_j^{t}\},\mathbf{y}^{t}\right)=0$, where $\mathbf{x}^{t+1}$ and $\mathbf{z}^{t+1}$ are the minimizers of (\ref{eq_18a}) and (\ref{eq_18b}), respectively, and  $\varsigma_{\mathbf{z}_j^{t+1}}$ is a sub-gradient vector. $\varsigma_{\mathbf{z}_j^{t+1}}$ implies $\mathcal{L}\left(\mathbf{x},\{\mathbf{z_j}\},\mathbf{y}\right)$ is non-differentiable function w.r.t. $\mathbf{z}_j$ can be observed in (\ref{eq_aug_DCMA}).
   
  By combining the inequalities (\ref{ap.2}) and (\ref{ap.3}), equation (\ref{eq_34}) can be written as
  \begin{align}\small
  \begin{split}
      \mathcal{L}\left(\mathbf{x}^{t+1},\{\mathbf{z}_j^{t+1}\},\mathbf{y}^{t+1}\right) - \mathcal{L}\left(\mathbf{x}^{t},\{\mathbf{z}_j^{t}\},\mathbf{y}^{t}\right) \\
      \leq -\frac{\gamma_1}{2}\Vert  \mathbf{x}^{t+1}-\mathbf{x}^{t} \Vert^2-\sum_{j=1}^{J }\frac{\gamma_2}{2}\Vert  \mathbf{z}^{t+1}-\mathbf{z}^{t} \Vert^2\\+ \frac{\lambda_{\rm max}^2(\mathbf{H}^H\mathbf{H})}{\rho} \Vert \mathbf{x}^{t+1}-\mathbf{x}^{t}\Vert^2\\
      \leq \left( \frac{\lambda_{\rm max}^2(\mathbf{H}^H\mathbf{H}) }{\rho}   -\frac{\gamma_1}{2}  \right)\Vert  \mathbf{x}^{t+1}-\mathbf{x}^{t} \Vert^2\\
      -\sum_{j=1}^{J}\frac{\gamma_2}{2}\Vert  \mathbf{z}^{t+1}-\mathbf{z}^{t} \Vert^2  
      \end{split}
  \end{align}
 If
    $\rho\gamma_1\geq 2\lambda_{\rm max}^2(\mathbf{H}^H\mathbf{H})$ and $\gamma_2> 0$,   then the augmented Lagrangian function always decreases. It is always possible to find a $\rho$ which satisfy the conditions $\rho\gamma_1\geq 2\lambda_{\rm max}^2(\mathbf{H}^H\mathbf{H})$ based on $\mathbf{H}$ along with $\gamma_2> 0$.
    \end{proof}
    \end{lemma}
    \begin{lemma}
        Let $\{\mathbf{x}^{t},\{\mathbf{z}^{t}\}_{j=1}^J,\mathbf{y}^{t}\}$ be generated by Algorithm~\ref{alg1}. Assume  $\rho\geq \lambda_{\rm max}(\mathbf{H}^H\mathbf{H})$. The following lower bound exists  
        \begin{align} \small          
        \mathcal{L}\left(\mathbf{x}^{t},\{\mathbf{z}_j^{t}\},\mathbf{y}^{t}\right) \geq
        f(\mathbf{z}^{t})+\alpha_1\sum_{j=1}^{J} w_j\Vert \mathbf{z}_j^{t}\Vert^2.
         \end{align}        
        \begin{proof}
        	Recall the augmented Lagrangian function  of sparse group LASSO formulated for DCMA detection
        \begin{align} \small 
        \begin{split}
            \mathcal{L}\left(\mathbf{x}^{t},\{\mathbf{z}_j^{t}\},\mathbf{y}^{t}\right) =
            f(\mathbf{x}^{t})+\alpha_1\sum_{j=1}^J w_j\Vert \mathbf{z}_j^{t}\Vert^2\\+ \innerproduct{\nabla_{\mathbf{x}}  f(\mathbf{x}^{t})}{\mathbf{z}^{t}-\mathbf{x}^{t}} + \frac{\rho}{2}\Vert\mathbf{x}^{t}-\mathbf{z}^{t}\Vert^2\label{aug_Lag.10}
            \end{split}
            \end{align} 
        The (\ref{aug_Lag.10}) uses the equality in (\ref{ap.01}).
            The  $\nabla_{\mathbf{x}}f(\mathbf{x})$ is Lipschitz continuous with constant $\lambda_{\rm max}(\mathbf{H}^H\mathbf{H})$ according to (\ref{ap.02}).   Thus, according to descent Lemma \cite{bertsekas2016nonlinear} the upper quadratic approximation for $f$ is given by  \vspace{-2mm}
            \begin{align*}
            \begin{split}
            f(\mathbf{z}^{t})\leq f(\mathbf{x}^{t})+\innerproduct{\nabla_{\mathbf{x}}  f(\mathbf{x}^{t})}{\mathbf{z}^{t}-\mathbf{x}^{t}}\\ + \frac{\lambda_{\rm max}(\mathbf{H}^H\mathbf{H})}{2}\Vert \mathbf{z}^{t}-\mathbf{x}^{t}\Vert^2   
            \end{split}
            \end{align*}
            which can be written as \vspace{-2mm}
            \begin{align}
            \begin{split}
            f(\mathbf{x}^{t})+ \innerproduct{\nabla_{\mathbf{x}}  f(\mathbf{x}^{t})}{\mathbf{z}^{t}-\mathbf{x}^{t}}\geq   f(\mathbf{z}^{t})\\-\frac{\lambda_{\rm max}(\mathbf{H}^H\mathbf{H})}{2}\Vert \mathbf{z}^{t}-\mathbf{x}^{t}\Vert^2  \label{ap.11}   
            \end{split}
            \end{align}
            Plugging (\ref{ap.11}) into (\ref{eq_aug_DCMA}), we get
            \begin{align*}\small
            \begin{split}
            \mathcal{L}\left(\mathbf{x}^{t},\{\mathbf{z}_j^{t}\},\mathbf{y}^{t}\right) \geq
            f(\mathbf{z}^{t})+\alpha_1\sum_{j=1}^J w_j\Vert \mathbf{z}_j^{t}\Vert^2\\+ \left( \frac{\rho}{2}-\frac{\lambda_{\rm max}(\mathbf{H}^H\mathbf{H})}{2}\right) \Vert \mathbf{z}^{t}-\mathbf{x}^{t}\Vert^2. \label{lem_3}
            \end{split}
            \end{align*}
           Suppose  $\rho\geq \lambda_{\rm max}(\mathbf{H}^H\mathbf{H})$ and $ f(\mathbf{z}^{t})+ \sum_{j=1}^{N\rm u} \frac{\gamma_j}{2} \Vert \mathbf{z}^{t}\Vert^2$ is bounded, due to the fact that $\mathbf{z}^{t}$ is belongs to finite energy codebooks. 
            Hence, $ \mathcal{L}\left(\mathbf{x}^{t},\{\mathbf{z}_j^{t}\},\mathbf{y}^{t}\right) \geq
            f(\mathbf{z}^{t})+ \sum_{j=1}^{N\rm u} \frac{\gamma_j}{2} \Vert \mathbf{z}^{t}\Vert^2$, lower bounded. This combined with (\ref{ap.1}) in \textbf{Lemma} 2, whenever the penalty parameter $\rho $ chosen sufficiently large then $ \mathcal{L}\left(\mathbf{x}^{t},\{\mathbf{z}_j^{t}\},\mathbf{y}^{t}\right)$ is monotonically decreasing and lower bounded. The two assumptions made in the beginning of the Appendix are validated.
            \end{proof}
              \end{lemma}            
    Now the proof for {\bf{Theorem-1}} is presented below.
    \begin{proof}
    According to Lemma-2, adding both sides of the (\ref{ap.1}) for $t=1,\hdots,\infty$, we get
    \begin{align}\small
    \begin{split}
    \mathcal{L}\left(\mathbf{x}^{1},\{\mathbf{z}_j^{1}\},\mathbf{y}^1\right)- \lim_{t\to\infty}\mathcal{L}\left(\mathbf{x}^{t},\{\mathbf{z}_j^{t}\},\mathbf{y}^{t}\right) \\
    \geq\left(   \frac{\rho+\lambda_{min}(\mathbf{H}^H\mathbf{H})}{2}   -\frac{\lambda_{\rm max}^2(\mathbf{H}^H\mathbf{H})}{\rho}     \right)\sum_{t=1}^{\infty}\Vert \mathbf{x}^{t+1}-\mathbf{x}^{t}  \Vert^2\\+\sum_{t=1}^{\infty} \frac{\rho}{2}\Vert \mathbf{z}^{t+1}-\mathbf{z}^{t}\Vert^2          \end{split}
    \end{align}
    From Lemma-3, $\mathcal{L}\left(\mathbf{x}^{t},\{\mathbf{z}_j^{t}\},\mathbf{y}^{t}\right) $ is lower bounded i.e. $\lim_{t\to\infty}\mathcal{L}\left(\mathbf{x}^{t},\{\mathbf{z}_j^{t}\},\mathbf{y}^{t}\right) > -\infty$. Furthermore, if $ \frac{\rho+\lambda_{min}(\mathbf{H}^H\mathbf{H})}{2} -\frac{\lambda_{\rm max}^2(\mathbf{H}^H\mathbf{H}) }{\rho}   >0$ and $\rho>0$,  the following limit can be obtained:
    \begin{align}\small\label{lim_1}
        \lim_{t\to\infty} \Vert \mathbf{x}^{t+1}-\mathbf{x}^{t}\Vert=0
    \end{align}
    \begin{align}\small\label{lim_2}
        \lim_{t\to\infty} \Vert \mathbf{z}^{t+1}-\mathbf{z}^{t}\Vert=0
    \end{align}
    From Lemma-1, $\Vert \mathbf{y}^{t+1}-\mathbf{y}^{t}\Vert\leq\lambda_{\rm max}(\mathbf{H}^H\mathbf{H})\Vert \mathbf{x}^{t+1}- \mathbf{x}^{t}\Vert,\forall j, k$
    \begin{align}\small
        \lim_{t\to\infty} \Vert \mathbf{y}^{t+1}-\mathbf{y}^{t}\Vert=0.
    \end{align}
    As,  $\mathbf{y}^{t+1}=\mathbf{y}^{t}+\rho(\mathbf{x}^{t+1}-\mathbf{z}^{t+1})$, the following limit exists
    \begin{align}
        \lim_{t\to\infty} \Vert \mathbf{x}^{t+1}-\mathbf{z}^{t+1}\Vert=0\label{ap.04}
    \end{align}
    Since the real and imaginary parts of $ \mathbf{x}$ and $ \mathbf{z}$ are bounded over $[\alpha_j, \alpha_j]$ and $[-\beta_j, \beta_j]$, respectively, then the limit point exists for $\mathbf{x}$ and $\mathbf{z}$ from (\ref{lim_1}) and (\ref{lim_2}) as follows
    \begin{align}
        \lim_{t\to\infty} \mathbf{x}^t=\mathbf{x}^\star,\quad
    \lim_{t\to\infty} \mathbf{z}^t=\mathbf{z}^\star\label{ap.03}
    \end{align}
    Plugging (\ref{ap.03}) into (\ref{ap.04}), then
    \begin{align}
        \lim_{t\to\infty} \mathbf{x}^t=\mathbf{x}^\star=\mathbf{z}^\star
    \end{align}
      Since $\mathbf{x}$ is bounded as mentioned above, from (\ref{ap.05}), the stationary point exists for $\mathbf{y}$ as follows
      \begin{align}
       \lim_{t\to\infty}  \mathbf{y}^{t}= \mathbf{y}^{\star}.         
      \end{align}
      Hence, $\{\mathbf{x}^{\star},\mathbf{z}^{\star},\mathbf{y}^{\star}\}_{j=1}^J$ is the stationary solution. 
            \end{proof}
        	\subsection{Convergence of ADMM algorithm on  problem~(\ref{eq_Case2b})}
        	The above derivations on convergence of ADMM on (\ref{eq_LASS_DCMA}) holds true for prior-information aided ADMM algorithm on~(\ref{eq_Case2b}).  The augmented Lagrangian function for the problem (\ref{eq_Case2b}) is given by

        	\begin{align}\small\label{aug_pri}
        		\begin{split}
        			\mathcal{L}\left(\mathbf{x}^{[l]},\{\mathbf{z}_j^{[l]}\}_{j=1}^J,\mathbf{y}_1^{[l]}\right)=	\frac{1}{2}\Vert \mathbf{r}^{[l]}-\mathbf{H}^{[l]}\mathbf{x}^{[l]}\Vert_2^2\\+\alpha_1\sum_{j=1}^Jw_{j,p}^l w_{j,d}\Vert\mathbf{z}_j\Vert_2 \frac{\mu^{l}}{2}\Vert \mathbf{x}^{[l]}-\hat{\beta}_{q^{[l]}}^{[l-1]} \Vert^2+ {\text{Re}}\innerproduct{\mathbf{y}_1^{[l]}}{\mathbf{x}^{[l]}-\mathbf{z}^{[l]}}\\
        			+\frac{\rho}{2}\Vert \mathbf{x}^{[l]}-\mathbf{z}^{[l]}\Vert_2^2
        		\end{split}
        	\end{align}
        In this case $f(\mathbf{x})$ in Subsection~\ref{sub_app} turns out to be $f(\mathbf{x}^{[l]})$ as given by $$f(\mathbf{x}^{[l]})=\frac{1}{2}\Vert \textbf{r}^{[l]}-\mathbf{H}^{[l]}\mathbf{x}^{[l]}\Vert_2^2+\frac{\mu^{l}}{2}\Vert \mathbf{x}^{[l]}-\hat{\beta}_{q^{[l]}}^{[l-1]} \Vert^2,$$
        By following the similar procedure in Subsection~\ref{sub_app}, the conditions on $\rho$ in problem (\ref{aug_pri}) are   $\rho(\rho+\mu^{l}+\lambda_{\rm min}(\mathbf{H}^H\mathbf{H}))\geq 2( \lambda_{\rm max}(\mathbf{H}^H\mathbf{H})+\mu^{l})^2$  and $\rho>~(\lambda_{\rm max}(\mathbf{H}^H\mathbf{H})+\mu^{l})$.

    


\end{document}